\documentclass[final,5p,times,twocolumn]{elsarticle}

\usepackage{amsmath}
\usepackage{amssymb}
\usepackage{bm}
\usepackage{siunitx}
\usepackage[version=4]{mhchem}
\usepackage{booktabs}
\usepackage{multirow}
\usepackage{array}
\usepackage{xcolor}
\usepackage{cleveref} 

\biboptions{sort&compress}

\journal{Journal of Magnetism and Magnetic Materials}

\begin{document}

\begin{frontmatter}

\title{Stripe antiferromagnetic order and spin dynamics in the Shastry-Sutherland Ising compound \ce{Er2Be2GeO7}}

\author[aff1,aff2,aff3]{Donger Cheng\fnref{equal}}
\author[aff4]{Otkur Omar\fnref{equal}}
\author[aff1,aff2]{Wenbin Wei}
\author[aff5]{Jan Peter Embs}
\author[aff1,aff2]{Xiao Wang}
\author[aff6]{Cheng Su}
\author[aff1,aff2]{Hui Cheng}
\author[aff7]{Zhengdong Fu}
\author[aff3]{Xinzhi Liu}
\author[aff6]{Wentao Jin}
\author[aff1,aff2]{Ze Hu}
\author[aff1,aff2]{Wei Luo}
\author[aff1,aff2]{Liming Wang}
\author[aff1,aff2]{Jiazheng Hao}
\author[aff1,aff2]{Feiran Shen}
\author[aff8,aff2]{Lunhua He}

\author[aff4]{Shang Gao\corref{cor1}}
\ead{sgao@ustc.edu.cn}

\author[aff1,aff2]{Erxi Feng\corref{cor1}}
\ead{fengex@ihep.ac.cn}

\fntext[equal]{Donger Cheng and Otkur Omar contributed equally to this work.}

\cortext[cor1]{Corresponding authors.}

\affiliation[aff1]{
    organization={Institute of High Energy Physics, Chinese Academy of Sciences (CAS)},
    city={Beijing},
    postcode={100049},
    country={China}
}

\affiliation[aff2]{
    organization={Spallation Neutron Source Science Center},
    city={Dongguan},
    postcode={523803},
    state={Guangdong},
    country={China}
}

\affiliation[aff3]{
    organization={Guangdong Provincial Key Laboratory of Magnetoelectric Physics and Devices, Center for Neutron Science and Technology, School of Physics, Sun Yat-sen University},
    city={Guangzhou},
    postcode={510275},
    state={Guangdong},
    country={China}
}

\affiliation[aff4]{
    organization={Department of Physics, University of Science and Technology of China},
    city={Hefei},
    postcode={230000},
    state={Anhui},
    country={China}
}

\affiliation[aff5]{
    organization={PSI Center for Neutron and Muon Sciences, Laboratory for Neutron Scattering and Imaging, Paul Scherrer Institut (PSI)},
    city={Villigen},
    postcode={CH-5232},
    country={Switzerland}
}

\affiliation[aff6]{
    organization={School of Physics, Beihang University},
    city={Beijing},
    postcode={100191},
    country={China}
}

\affiliation[aff7]{
    organization={Songshan Lake Materials Laboratory},
    city={Dongguan},
    postcode={523808},
    country={China}
}

\affiliation[aff8]{
    organization={Institute of Physics, Chinese Academy of Sciences (CAS)},
    city={Beijing},
    postcode={100049},
    country={China}
}

\begin{abstract}
In rare-earth compounds, the interplay between geometric frustration and unusual
single-ion anisotropy caused by the crystal electric field often leads to exotic
magnetic phases. Here, we investigate the magnetic properties of the rare-earth
Shastry-Sutherland compound \ce{Er2Be2GeO7} using bulk measurements and neutron
scattering down to sub-Kelvin temperatures. A $\boldsymbol{k}=(0,0,0)$ stripe-like antiferromagnetic order with locally tilted Er$^{3+}$ moments is revealed by
time-of-flight neutron powder diffraction, accompanied with a $\lambda$-shape
specific-heat anomaly at $T_\mathrm{N}=0.84$~K. The temperature dependence of the magnetic peak intensity corresponding to the ordered moment of Er$^{3+}$ approximates to the two-dimensional Ising universality class. Thermal- and cold-neutron inelastic scattering experiments demonstrate that the strong Ising anisotropy originates from the local crystal electric field, and reveal gapped magnetic excitations with weak dispersion in the ordered state. Mean-field random-phase approximation with inequivalent nearest-neighbor couplings accounts well for the collective excitations, while classical Monte Carlo simulations yield the fractional $1/4$ and $1/2$ magnetization plateaus. Our work provides a comprehensive description of the spin correlations in \ce{Er2Be2GeO7} with a particular focus on the classical origin of its fractional magnetization plateaus.
\end{abstract}

\begin{keyword}
Frustrated magnetism
\sep Shastry-Sutherland lattice
\sep Collective magnetic excitations
\end{keyword}

\end{frontmatter}


\section{Introduction}
\label{sec:introduction}

Frustrated magnets provide a fertile ground for unconventional magnetic orders and novel quantum phases arising from competing exchange interactions~\cite{balentsSpinLiquidsFrustrated2010,broholmQuantumSpinLiquids2020a}. The Shastry-Sutherland lattice (SSL) is a unique exactly solvable model, where magnetic ions form an orthogonal network of dimers and are connected by intra-dimer ($J_1$) and inter-dimer ($J_2$) exchange interactions~\cite{sriramshastryExactGroundState1981a}. The coupling ratio $J_1/J_2$ determines a rich phase diagram, which is realized in the $S = 1/2$ compound \ce{SrCu2(BO3)2}~\cite{knetterDispersionSymmetryBound2000,kageyamaExactDimerGround1999c,mullerhartmannExactDemonstrationMagnetization2000,kageyamaCrystalGrowthTwodimensional1999,kageyamaDirectEvidenceLocalized2000,kodamaMagneticSuperstructureTwoDimensional2002a,lemmensCollectiveSingletExcitations2000,nomuraUnveilingNewQuantum2023,zayed4spinPlaquetteSinglet2017,shiDiscoveryQuantumPhases2022}. Incorporating rare-earth (RE) ions into the SSL introduces strong single-ion anisotropy due to the crystal electric field (CEF) and profoundly enriches the magnetic phase diagram. Unlike metallic systems where Ruderman-Kittel-Kasuya-Yosida (RKKY) interactions obscure the intrinsic SSL physics~\cite{kim$textYb_2textPt_2textPb$MagneticFrustration2008,kimSpinLiquidsAntiferromagnetic2013,matsumuraNonCollinearMagneticStructure2007a,metokiMagneticQuadrupoleCoupling2025}, the insulating melilite family \ce{RE2Be2GeO7} provides an ideal platform to probe intrinsic magnetic behaviors of the SSL model~\cite{ashtarStructureMagneticProperties2021b,liuLargeMagnetocaloricEffect2024,liuDistinctMagneticGround2024a}. 

Among the melilite family, \ce{Er2Be2GeO7} has recently attracted attention for its unconventional magnetization behavior. Pula \cite{Pula_2025} and Yadav~\cite{yadavObservationUnprecedentedFractional2025a} independently reported the emergence of fractional $1/4$ and $1/2$ magnetization plateaus below $0.5 \,\mathrm{T}$, a notable deviation from the expected $1/3$ plateau of the SSL Ising model~\cite{dublenych_ground_2012}. Single-crystal neutron diffraction further revealed a subtle structural distortion from space group $P\overline{4}2_1m$ to $Cmm2$~\cite{yadavObservationUnprecedentedFractional2025a}, which has been proposed to underlie the unusual magnetization plateaus. Previous theoretical work reproduced the magnetization plateaus in \mbox{\ce{Er2Be2GeO7}} using a classical Ising Shastry-Sutherland model via the corner transfer matrix renormalization group (CTMRG) method~\mbox{\cite{yadavObservationUnprecedentedFractional2025a}}, thus suggesting the classical origin of the magnetization plateaus like that established in ErB$_4$~\mbox{\cite{huang_multi_2012, dublenych_ground_2012}}. However, it is unclear whether such a classical interpretation could be reconciled with the spin dynamics in \mbox{\ce{Er2Be2GeO7}}.

Here, through combined neutron scattering experiments and numerical simulations on the spin dynamics, we propose an Ising Shastry-Sutherland model for \ce{Er2Be2GeO7} and confirm that its magnetization plateaus are essentially classical behavior governed by strong Ising anisotropy. Our neutron diffraction determines a commensurate non-collinear magnetic order with moments tilted away from the $c$-axis, consistent with the previous report~\cite{yadavObservationUnprecedentedFractional2025a}. Inelastic neutron scattering (INS) reveals gapped collective magnetic excitations with weak dispersion. Through mean-field random-phase-approximation (MF-RPA) analysis, we propose an Ising Shastry-Sutherland model that quantitatively reproduces the INS spectra. This model agrees with the previous theoretical analysis in its incorporation of the 
orthorhombic lattice distortion~\cite{yadavObservationUnprecedentedFractional2025a}. Using the same model, magnetization plateaus in \ce{Er2Be2GeO7} can be reproduced via classical Monte Carlo simulations, thus confirming classical Ising behavior as the main origin.

\section{Experimental details}
\label{sec:experimental}
Polycrystalline \(\ce{Er2Be2GeO7}\) sample was synthesized via solid-state reaction method~\cite{ashtarStructureMagneticProperties2021b}. Phase purity and crystal structure were confirmed by X-ray diffraction (XRD) using a Rigaku diffractometer equipped with Cu K$_\alpha$ radiation ($\lambda = \SI{1.5418}{\angstrom}$) over the angular range of $2\theta = 5 - \SI{90}{\degree}$. Time-of-flight neutron powder diffraction (NPD) were performed on the General Purpose Powder Diffractometer (GPPD) at the China Spallation Neutron Source (CSNS)~\cite{gppd} to probe the magnetic order. Around \SI{3}{\gram} powder sample was sealed in a Cu hollow cylinder sample can with a thickness of \SI{2}{\mm} in He gas. Diffraction patterns were collected with two detector banks located at scattering angles of \(2\theta = \SI{30}{\degree}\) and \(\SI{90}{\degree}\). Rietveld refinements for the XRD and NPD data were performed using the FULLPROF suite
\cite{mccuskerRietveldRefinementGuidelines1999}.

Magnetic susceptibility and magnetization were measured using a Quantum Design Physical Property Measurement System (PPMS) equipped with a Vibrating Sample Magnetometer (VSM) module. Temperature-dependent magnetic susceptibility and isothermal magnetization measurements were performed between \SI{2} and \SI{300}{\kelvin}. Specific heat measurements on \ce{Er2Be2GeO7} and the non-magnetic isostructural reference compound \ce{Lu2Be2GeO7} were carried out using the heat capacity module of the PPMS.  Ultralow-temperature specific heat data down to \SI{60}{\milli\kelvin} under magnetic fields were obtained using a dilution refrigerator insert.

Inelastic neutron scattering (INS) experiments were employed to observe the CEF excitations and the low-lying collective magnetic excitations. Thermal neutron INS measurements were carried out on the High-Energy Direct Geometry Spectrometer (HD)~\cite{luoDesignHighEnergy2023a} at the CSNS with incident energies $E_i$ of $10, 30$, and \SI{60}{\milli\electronvolt} at \SI{5}{\kelvin}. Cold INS measurements were conducted on the time-of-flight spectrometer FOCUS at the Paul Scherrer Institute (PSI)~\cite{FOUCS1996}, with an incident energy of $E_i$ = \SI{3.55}{\milli\electronvolt} at temperatures of $0.05, 0.3, 0.8$, and \SI{1.89}{\kelvin}. The INS data were reduced and analyzed using the DAVE/MSlice software package~\cite{azuahDAVEComprehensiveSoftware2009}.

\section{Results and Discussion}
\label{sec:results}
\subsection{Crystal  structure}

\begin{figure}[h!]
    \centering
        \includegraphics[width=\columnwidth]{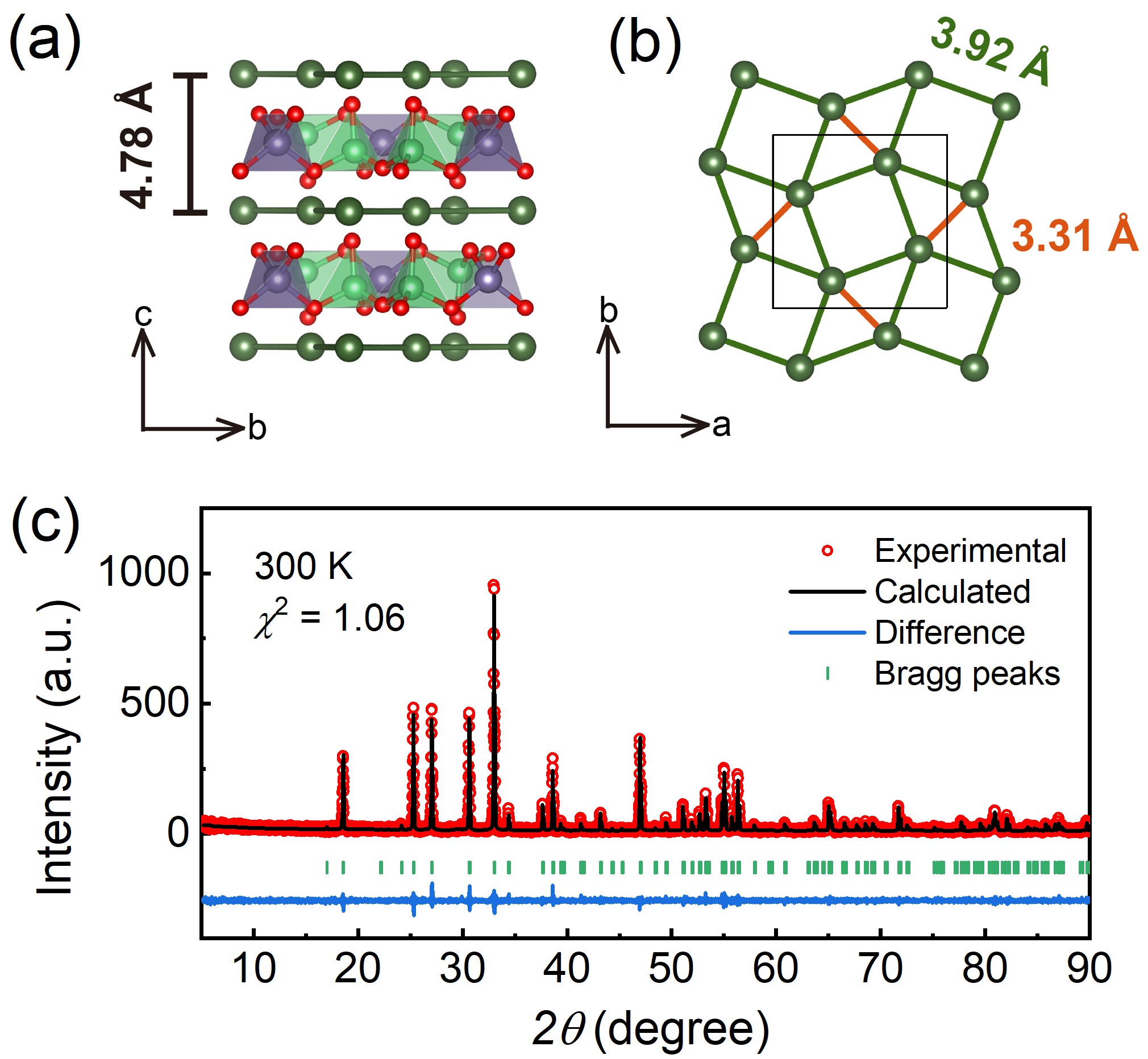}
        \caption{(a) Crystal structure of \ce{Er2Be2GeO7} viewed along the $c$-axis, showing magnetic ion layers separated by nonmagnetic polyhedra with the space of \SI{4.78(2)}{\angstrom}. (b) The SSL motif formed by the \ce{Er^{3+}} ions in the $ab$-plane. The intradimer and interdimer bond lengths are \SI{3.31(6)}{\angstrom} and \SI{3.92(4)}{\angstrom}, respectively. (c) Rietveld refinements of the powder XRD pattern at room temperature. Red circles, black line and blue line are the experimental, calculated intensities and difference, respectively.  Green ticks represent the positions of the Bragg reflections.}
        \label{Fig: Figure1}
\end{figure}

\ce{Er2Be2GeO7} crystallizes in the tetragonal space group $P\overline{4}2_1m$ (No. 113), as illustrated in Fig.~\ref{Fig: Figure1}(a), the magnetic \ce{Er^{3+}} layers separated by nonmagnetic \ce{[GeBe2O7]^{6-}} polyhedra along the $c$-axis. Within the $ab$-plane, the \ce{Er^{3+}} ions form a SSL with AA-type stacking along the $c$-axis [Fig.~\ref{Fig: Figure1}(b)]. The large interlayer spacing of $c/2$ suggests weak interlayer magnetic interactions, rendering the system quasi-two-dimensional. The in-plane intradimer and interdimer Er--Er distances are  \SI{3.31(6)}{\angstrom} and \SI{3.92(4)}{\angstrom}, respectively, corresponding to the exchange interactions $J_1$ and $J_2$ in the SSL model. Phase purity and room-temperature structural parameters were determined by Rietveld refinement of XRD data [Fig.~\ref{Fig: Figure1}(c)], yielding $\chi^{2} = 1.06$ and confirming the high quality of the sample.
A subtle orthorhombic distortion with space group $Cmm2$ was reported via single-crystal neutron diffraction ~\cite{yadavObservationUnprecedentedFractional2025a}, which leads to inequivalent intradimer couplings, $J_{1u} \neq J_{1v}$, as shown in Fig.~\ref{Fig: Figure3}(c). Unfortunately, this distortion could not be discerned in our powder neutron diffraction experiment due to the limitation of the $Q$-resolution of the diffractometer ($\Delta Q / Q \geq 0.2 \%$)~\cite{gppd}. Thus, the tetragonal structure was used as the parent structure for magnetic structure determination. 

\subsection{Magnetic properties}
\begin{figure}[t]
		\centering
		\includegraphics[width=\linewidth]{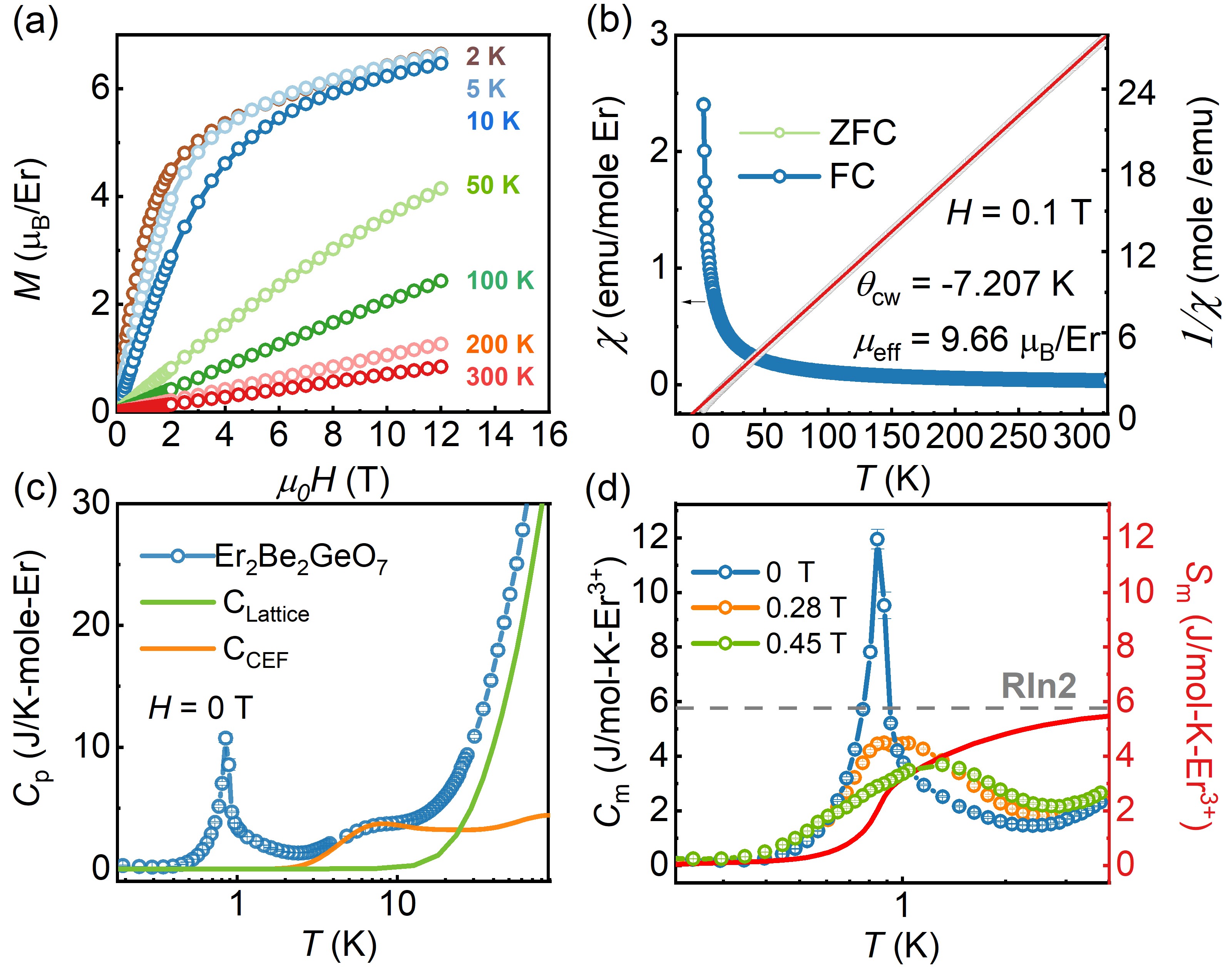}
		\caption{(a) Isothermal $M(H)$ at representative temperatures. (b) Magnetic susceptibility $\chi(T)$ and inverse susceptibility $1/\chi(T)$ at \SI{0.1}{\tesla}; red line: Curie--Weiss fit. (c) Zero-field specific heat $C_p$. (d) Magnetic specific heat $C_m$ at selected fields and zero-field magnetic entropy $S_{mag}$.}
		\label{Figure2}
	\end{figure}

Isothermal magnetization curves measured at various temperatures are presented in Fig.~\ref{Figure2}(a). The magnetization at \SI{2}{\kelvin} and \SI{12}{\tesla} reaches approximately $6.8\,\mu_{\mathrm{B}}$ per \ce{Er^{3+}}, which is lower than the theoretical saturation value of $9\,\mu_{\mathrm{B}}$ expected for the fully polarized ($J = 15/2$) state. This lower value results from the strong Ising-like anisotropy due to the CEF, which prevents the randomly oriented magnetic moments in the powder sample from being fully aligned by the external field.

Fig.~\ref{Figure2}(b) shows the temperature dependence of the magnetic susceptibility $\chi(T)$ and inverse susceptibility $1/\chi(T)$ measured at \SI{0.1}{\tesla}. No magnetic ordering transition is observed in the temperature range from $2$ to \SI{300}{\kelvin}. The susceptibility follows the Curie-Weiss law $\chi = C/(T - \theta_{\text{CW}})$ in the high-temperature regime $150$ to \SI{300}{\kelvin}, yielding a Curie-Weiss temperature $\theta_{\text{CW}} = \SI{-7.21(4)}{\kelvin}$ and an effective magnetic moment $\mu_\mathrm{eff}=9.664(1) ~\mu_\mathrm{B}$ per \ce{Er^{3+}} ion, calculated from the Curie constant $C$ using ${\mu_{\text{eff}} = \sqrt{3k_{\mathrm{B}} C/N_{\mathrm{A}}}}$. This value is in excellent agreement with the theoretical free-ion value of $9.6\,\mu_\mathrm{B}$ for \ce{Er^{3+}} (${}^4I_{15/2}$). The negative $\theta_{\text{CW}}$ indicates dominant antiferromagnetic interactions, while the small magnitude suggests competing exchange interactions. The frustration parameter $f = |\theta_{\text{CW}}|/T_{\text{N}} \approx 8.58(5)$ (where $T_{\text{N}} = \SI{0.84}{\kelvin}$ from specific heat measurements below) indicates moderate strength of the magnetic frustration in this system.

Thermodynamic measurements confirm the magnetic transition. Fig.~\ref{Figure2}(c) shows the specific heat $C_{\text{p}}$ for \ce{Er2Be2GeO7}, with a sharp $\lambda$-anomaly at $T_{\text{N}} = \SI{0.84}{\kelvin}$ indicating antiferromagnetic ordering and a broad Schottky anomaly near \SI{7.5}{\kelvin} from CEF excitations. The crystal electric field (CEF) contribution is modeled using a Schottky multilevel formula~\cite{schottky}:
\begin{equation}
\begin{aligned}
    C_{\text{sch}} &= \frac{R}{Z^2} \left[ Z \sum_{i=1}^{n} \left( \frac{E_i}{k_\mathrm{B}T} \right)^2 e^{-E_i / (k_\mathrm{B}T)}\right. \\
    &\quad \left. - \left\{ \sum_{i=1}^{n} \left( \frac{E_i}{k_\mathrm{B}T} \right) e^{-E_i / (k_\mathrm{B}T)} \right\}^2 \right],
\end{aligned}
\end{equation}
where $Z = \sum_{i=1}^{n} e^{-E_i/(k_B T)}$ is the partition function, $R$ is the universal gas constant, and $E_i$ are the CEF excitation energies determined from INS measurements. After subtracting the lattice contribution (from nonmagnetic \ce{Lu2Be2GeO7}) and CEF contribution~\cite{schottky}, the magnetic entropy $S_{\text{mag}}$ recovers \SI{5.60}{\joule\per\kelvin\per\mole} ($0.97\, R\ln2$) at \SI{4}{\kelvin}, consistent with a Kramers doublet ground state and confirming the effective spin-$1/2$ description. Under applied magnetic fields, the antiferromagnetic transition is rapidly suppressed. In contrast to previous single-crystal measurements, where a thermodynamic anomaly was observed at the $1/4$ plateau~\cite{yadavObservationUnprecedentedFractional2025a}, our powder data show only a broad maximum that becomes discernible at \SI{0.45}{\tesla}, corresponding to the $1/2$ plateau. This difference likely reflects the random orientation of crystallites in the powder sample, which averages out anisotropic field responses. These broad maxima nevertheless provide thermodynamic evidence for field-induced modifications of the magnetic ground state consistent with the stabilization of fractional plateau phases.

\subsection{Magnetic structure}
\begin{figure}[h]
    \centering
        \includegraphics[width=\columnwidth]{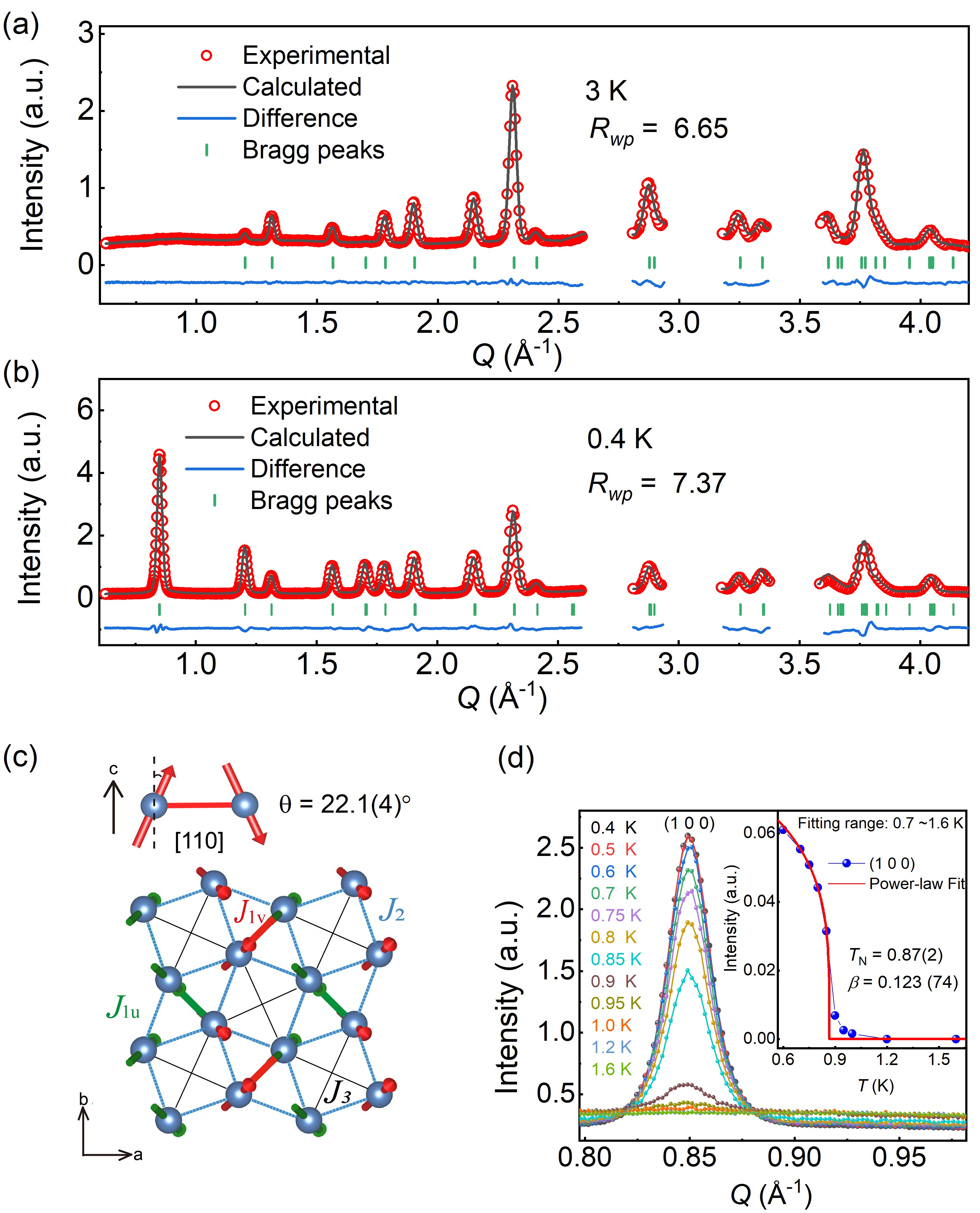}
        \caption{NPD patterns of $\ce{Er2Be2GeO7}$ collected at (a) \SI{3}{\kelvin } and (b) \SI{0.4}{\kelvin}. Data in the $Q$ ranges of $2.58$ to \SI{2.81}{\per\angstrom} and $2.93$ to \SI{3.19}{\per\angstrom} are excluded due to the background from the Cu holder and Al can. (c) schematic diagram of the spin arrangement with magnetic interactions indicated by color. $J_{1v}$ (red) and $J_{1u}$ (green) represent the inequivalent intradimer bonds resulting from orthorhombic distortion. $J_2$ (blue) and $J_3$ (black) denote two geometrically distinct inter-dimer exchange pathways. (d) Temperature evolution of the $(100)$ magnetic Bragg peak upon warming. Inset: temperature dependence of integrated intensity, with error bars smaller than the symbol size. The red solid line represents a power-law fit.}
        \label{Fig: Figure3}
\end{figure}

Time-of-flight neutron powder diffraction (NPD) patterns collected at $T=3$ and \SI{0.4}{\kelvin} are compared in Figs.~\ref{Fig: Figure3}(a) and (b). Below $T_{\text{N}} \approx \SI{0.87}{\kelvin}$, additional Bragg peaks emerge as magnetic long range ordering. Symmetry analysis identifies the magnetic space group  $P2_12_12'$ with a propagation vector $\boldsymbol{k} = (0, 0, 0)$. The refined ordered moment is $6.79(1)\,\mu_{\mathrm{B}}$ per \ce{Er^{3+}}. The spin configuration consists of nearly antiferromagnetic dimers forming a stripe-like arrangement in the $ab$-plane, while spins are ferromagnetically aligned along the $c$-axis. This configuration agrees well with the magnetic structure previously reported by Yadav et al.~\cite{yadavObservationUnprecedentedFractional2025a}. The ordered moments are canted by an angle $\theta = \SI{22.1(4)}{\degree}$ from the crystallographic $c$-axis, reflecting a locally tilted Ising easy axis primarily arising from the local CEF, as discussed below. The magnetic ordering behavior was characterized by the temperature evolution of the magnetic reflection peak $(100)$, as presented in Fig.~\ref{Fig: Figure3}(d). 
 The magnetic peak intensity increases below $T_{\text{N}}$, and its temperature dependence follows the $I(T) - I_0 \propto \left(1 - T/T_{\text{N}}\right)^{2\beta}$, yielding a transition temperature of $\SI{ 0.87(2)}{\kelvin}$ and $\beta = 0.123(74)$, in the critical temperature range near transition temperature. The obtained $\beta$ value is close to the theoretical two-dimensional Ising value. This similarity agrees with the Ising-type single-ion anisotropy of Er$^{3+}$ from the CEF analysis below ~\cite{yadavObservationUnprecedentedFractional2025a}.

\subsection{Single-ion anisotropy}
\begin{figure}[h]
    \centering
        \includegraphics[width=\columnwidth]{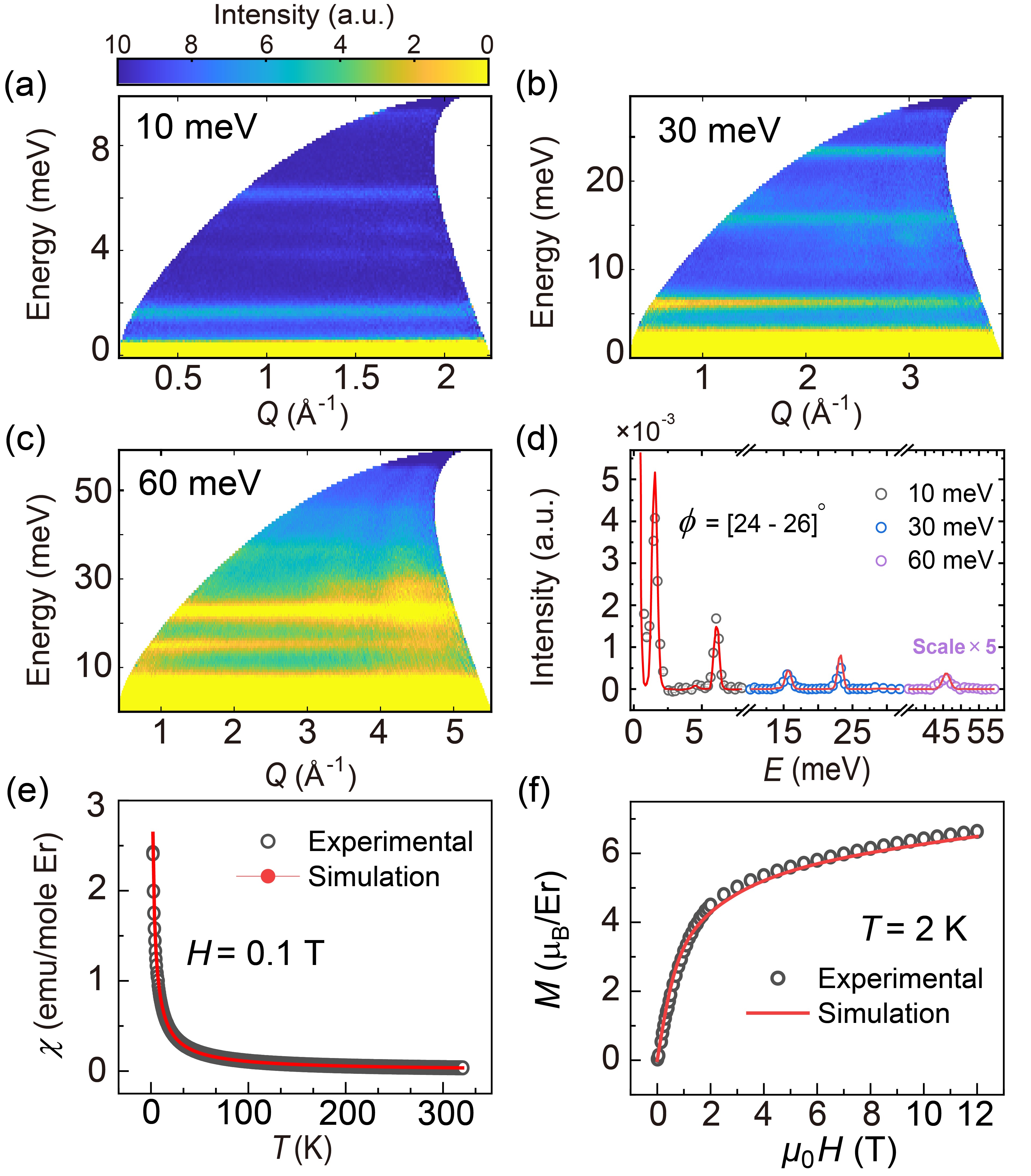}
        \caption{(a-c) INS spectra at \SI{5}{\kelvin} for $E_i=\,$10, 30, and \SI{60}{\milli\electronvolt} showing dispersionless CEF excitations. (d) INS intensities as a function of energy transfer, integrated between scattering angles \SI{24}{\degree} to \SI{26}{\degree} for the three $E_{i}$. The solid line represents the CEF fits to the experimental data. Magnetic susceptibility (e) and magnetization (f) of \ce{Er2Be2GeO7}. Symbols represent the experimental data, while the solid line shows the calculated results based on the best fitted CEF parameters listed in Table~\ref{tab:CEF_parameters}. The good agreement confirms the reliability of the refined crystal-field scheme.}
        \label{Fig: Figure4}
\end{figure}

To understand the single-ion magnetism, we performed thermal INS experiments to probe its CEF excitations ~\cite{luoDesignHighEnergy2023a}. As a Kramers ion with $4f^{11}$ configuration, the ground-state multiplet ${}^4I_{15/2}$ of \ce{Er^{3+}} is split by the CEF into eight doublets. In the INS spectra [ Figs.~\ref{Fig: Figure4}(a-c)], four excitations are well resolved at approximately \SI{1.57}, \SI{6.17}, \SI{15.79}, and \SI{23.30}{\milli\electronvolt}, with two weak INS peaks at higher energies (\SI{28.46} and \SI{45.27}{\milli\electronvolt}). 
The CEF acting around the \ce{Er^{3+}} ions arises from the electrostatic potential of the surrounding ligands. In \ce{Er2Be2GeO7}, each \ce{Er^{3+}} ion is coordinated by eight oxygen ions forming a distorted polyhedron. Crystallographic analysis indicates that the local environment possesses a $C_s$ point symmetry, characterized by a single mirror plane. The CEF Hamiltonian can be expressed using the tensor operator equivalents~\cite{hutchingsPointChargeCalculationsEnergy1964}:
\begin{equation}
\mathcal{H}_{\mathrm{CEF}} = \sum_{k=2,4,6} \sum_{q=-k}^{k} B_k^q O_k^q,
\label{eq:CEF_general}
\end{equation}
where $B_k^q$ are the CEF parameters and $O_k^q$ are the Stevens operators. The number of independent parameters depends critically on the point symmetry and the choice of coordinate system. For a site with a $C_s$ symmetry, the appropriate choice of coordinate axes significantly simplifies the Hamiltonian. If the coordinate system is chosen such that the $y$-axis is perpendicular to the mirror plane and the $xz$-plane lies within the mirror (that is the \{1 1 0\} planes in \ce{Er2Be2GeO7}), this convention requires all terms odd under the mirror operation to vanish, leading to $B_k^q = 0$ for all $q<0$. 
The Hamiltonian is therefore reduced to the following 15 independent terms: 

\begin{align}
\mathcal{H}_{\mathrm{CEF}} 
&= B_2^0 O_2^0 + B_2^1 O_2^1 + B_2^2 O_2^2 \notag \\
&\quad + B_4^0 O_4^0 + B_4^1 O_4^1 + B_4^2 O_4^2 + B_4^3 O_4^3 + B_4^4 O_4^4 \notag \\
&\quad + B_6^0 O_6^0 + B_6^1 O_6^1 + B_6^2 O_6^2 + B_6^3 O_6^3 + B_6^4 O_6^4 \notag \\
&\quad + B_6^5 O_6^5 + B_6^6 O_6^6, \label{eq:CEF}
\end{align}

Owing to the limited number of resolved excitations, the INS data are insufficient to uniquely determine all CEF parameters. We therefore used the reported CEF parameters ~\cite{yadavObservationUnprecedentedFractional2025a} as the initial model and refined them by considering both INS spectra and magnetic characterization. The resulting optimized parameters provide an improved overall description of both the excitation spectrum and the bulk magnetic behaviors. Table~\ref{tab:CEF_parameters} summarizes the comparison between the CEF parameters reported in Ref.~\cite{yadavObservationUnprecedentedFractional2025a} and the optimized values obtained in this work. The validity of our refined CEF parameters is strongly supported by the agreement between the measured and calculated data of the susceptibility and magnetization at 2 K, as shown in Fig.~\ref{Fig: Figure4}(e,f).
\begin{table}[h]
\caption{Comparison of CEF parameters for \ce{Er^{3+}} in \ce{Er2Be2GeO7}. The second column lists the parameters reported in Ref.~\cite{yadavObservationUnprecedentedFractional2025a}, while the third column shows the best-fitted values obtained in this work. All parameters are given in meV.}
\label{tab:CEF_parameters}
\centering
\begin{tabular}{ccc}
\hline\hline
Parameter & Ref.~\cite{yadavObservationUnprecedentedFractional2025a} & This work \\
\hline
$B_2^0$ & $-2.6110\times10^{-3}$ & $-2.7157\times10^{-3}$ \\
$B_2^1$ & $-4.7360\times10^{-1}$ & $-5.3528\times10^{-1}$ \\
$B_2^2$ & $\phantom{-}1.3250\times10^{-1}$ & $\phantom{-}1.7225\times10^{-1}$ \\
$B_4^0$ & $-6.7000\times10^{-5}$ & $-8.7089\times10^{-4}$ \\
$B_4^1$ & $\phantom{-}2.2730\times10^{-3}$ & $\phantom{-}2.6045\times10^{-3}$ \\
$B_4^2$ & $-2.9880\times10^{-3}$ & $-2.3976\times10^{-3}$ \\
$B_4^3$ & $-1.3240\times10^{-3}$ & $-1.4243\times10^{-3}$ \\
$B_4^4$ & $-5.7425\times10^{-4}$ & $-4.3147\times10^{-4}$ \\
$B_6^0$ & $-8.8900\times10^{-7}$ & $-7.1638\times10^{-7}$ \\
$B_6^1$ & $\phantom{-}1.1010\times10^{-4}$ & $\phantom{-}1.2581\times10^{-4}$ \\
$B_6^2$ & $-6.0660\times10^{-5}$ & $-4.4269\times10^{-5}$ \\
$B_6^3$ & $\phantom{-}7.8180\times10^{-5}$ & $\phantom{-}7.1842\times10^{-5}$ \\
$B_6^4$ & $\phantom{-}5.5700\times10^{-6}$ & $\phantom{-}6.3688\times10^{-6}$ \\
$B_6^5$ & $\phantom{-}7.5470\times10^{-5}$ & $\phantom{-}8.6338\times10^{-5}$ \\
$B_6^6$ & $\phantom{-}3.2620\times10^{-6}$ & $\phantom{-}4.2244\times10^{-6}$ \\
\hline\hline
\end{tabular}
\end{table}

\begin{figure}[t]
    \centering
    \includegraphics[width=\columnwidth]{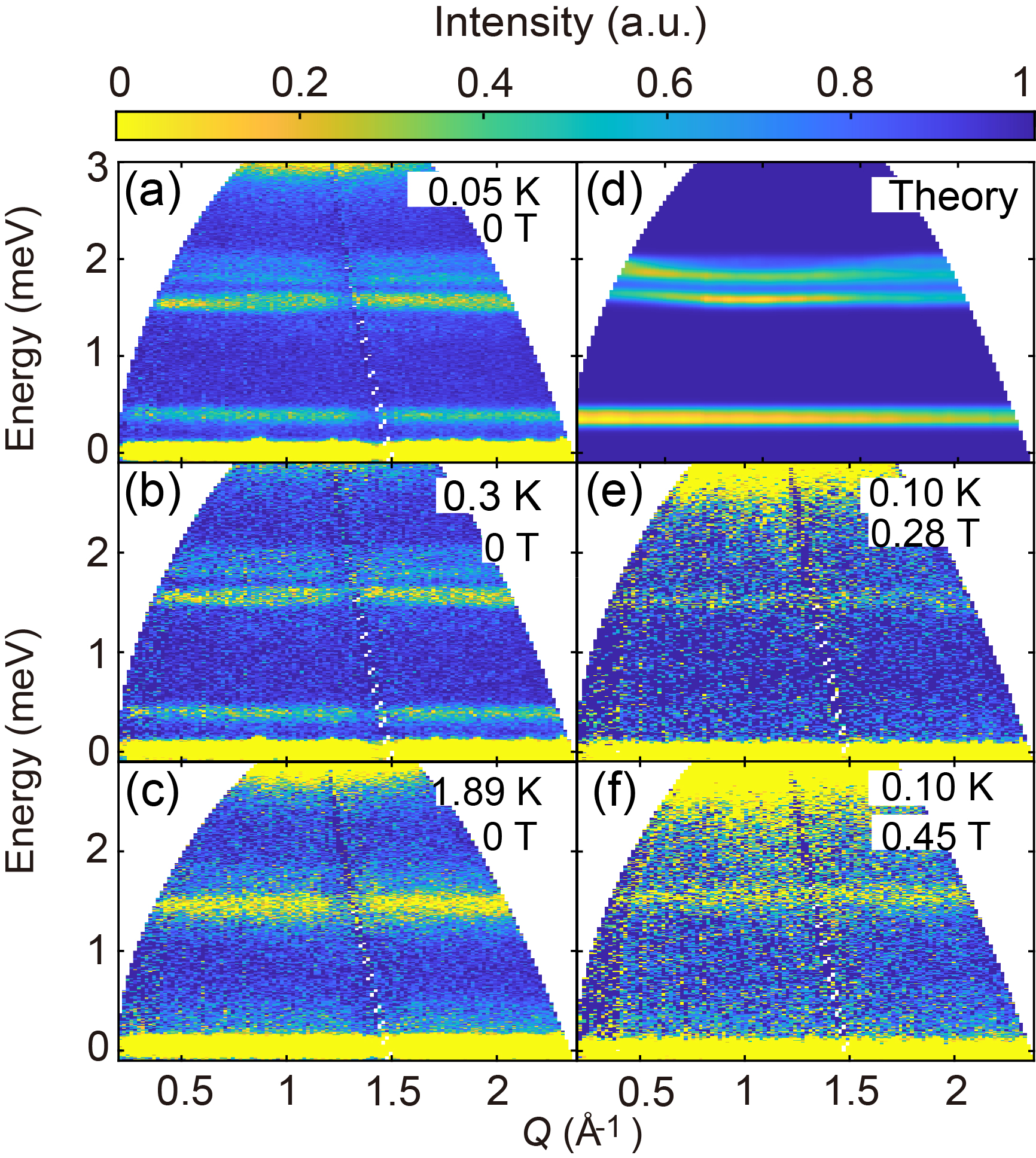}
        \caption{INS spectra measured on the cold neutron spectrometer FOCUS at PSI with an incident neutron energy $E_{\mathrm{i}}$ = \SI{3.55}{\milli\electronvolt}. Experimental (a) and  calculated (d) INS powder spectra of \ce{Er2Be2GeO7} at $T = \SI{0.05}{\kelvin}$ without magnetic field. (b, c) Zero field INS spectra measured at $T = 0.3$ and \SI{1.89}{\kelvin}, respectively. (e, f) INS spectra measured under magnetic fields of $\mu_{0}H = 0.28$ and \SI{0.45}{\tesla} at $T = \SI{0.1}{\kelvin}$, respectively.} 
        \label{Fig: Figure5}
\end{figure}

\subsection{Collective magnetic excitations}
Cold INS spectra at $T = 0.05$~K [Fig.~\ref{Fig: Figure5}(a)] show three nearly dispersionless magnetic excitations at $0.37$, $1.83$, and \SI{1.98}{\milli\electronvolt}, beside the first CEF transition at \SI{1.57}{\milli\electronvolt}. These modes exhibit temperature and field dependence, as illustrated in Fig.~\ref{Fig: Figure5}: they vanish above $T_{\text{N}}$ and are suppressed at the magnetic fields where $1/4$ and $1/2$ plateaus occur ~\cite{yadavObservationUnprecedentedFractional2025a}.
\begin{figure*}
    \centering
    \includegraphics[width=\textwidth]{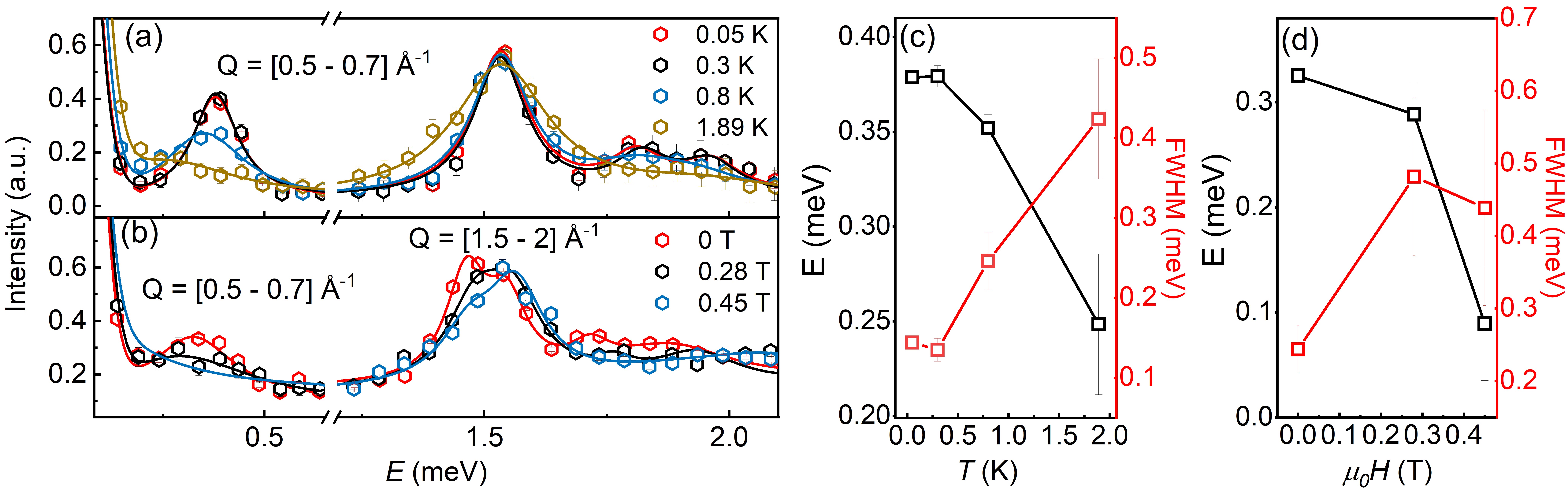}
        \caption{(a) Constant-$Q$ cut of the cold INS spectra at different temperatures under zero field, and (b) under various magnetic fields at $T = \SI{0.1}{\kelvin}$. The solid lines represent the total fits of the multi-component DHO function. (c,d) Temperature and magnetic field dependence of the gapped excitation at $E = \SI{0.37}{\milli\electronvolt}$, respectively. Black symbols denote the peak positions, and red symbols correspond to the linewidths extracted from the multi-component DHO fits.}
        \label{Fig: Figure6}
\end{figure*} To quantify the evolution of these excitations, we analyzed the momentum-integrated spectra using a Gaussian function for the elastic scattering contribution together with damped harmonic oscillator (DHO) functions for the inelastic magnetic excitations. The \SI{0.37}{\milli\electronvolt} mode exhibits typical gapped magnon behavior, as shown in Fig.~\ref{Fig: Figure6}(a,c): as temperature approaches $T_{\text{N}}$, the excitation softens with broadening linewidth. The behavior reflects the critical slowing down~\cite{roscildeScalableSpinSqueezing2024} of spin dynamics accompanying the diminishing of the long-range magnetic order. The two high-energy modes at $1.83$ and \SI{1.98}{\milli\electronvolt} gradually weaken with increasing temperature, while the fitted peak positions suggest a possible shift toward lower energies. Further insight into the origin of these excitations can be obtained from their magnetic-field dependence.

Under applied fields at $T = \SI{0.1}{\kelvin}$ [Figs.~\ref{Fig: Figure6}(b,d)],  the low- and higher-energy modes exhibit distinct responses. The low-energy mode at \SI{0.37}{\milli\electronvolt} softens and broadens continuously with increasing field, reflecting the destabilization of the stripe-type ordered state under magnetic field. In contrast, the higher-energy modes exhibit pronounced reductions in intensity and shifts toward higher energies. Such behavior is consistent with a single-ion origin, where the applied field induces Zeeman splitting of the CEF levels and alters their thermal population.
\subsection{Theoretical modeling}
\label{sec:simulations}

We employ a theoretical framework based on the mean-field random phase approximation (MF-RPA) to quantitatively analyze the magnetic ground state and excitations in \ce{Er2Be2GeO7}~\cite{Jensen:RareEarth,Rotter_2012,gwph-w9zm,PhysRevB.106.054431}. Compared to the effective spin-$\frac{1}{2}$ Ising model projected in the ground state doublet~\cite{yadavObservationUnprecedentedFractional2025a}, the MF-RPA framework is based on the full single-ion CEF Hamiltonian $H_\mathrm{CEF} = \sum_{k,q}B_k^qO_k^q$ without projection, which implicitly incorporates quantum tunneling that is required for spin dynamics in a longitudinal molecular field. The spin Hamiltonian, which considers only the longitudinal exchange field as the first approximation, is defined as:
\begin{equation}
\mathcal{H}=\sum_{k,q}B_k^qO_k^q(\tilde{\mathbf{J}})+\frac12\sum_{i\neq j}{\mathcal{J}}(i,j)\tilde{J}^i_{\tilde{z}}\tilde{J}_{\tilde{z}}^j,
\end{equation}
where $\tilde{\mathbf{J}}$ denotes the total angular momentum in the local coordinate frame, and $\tilde{J}_{\tilde{z}}$ is the component along the local Ising axis $\tilde{z}$ of each magnetic ion. The exchange interactions $\mathcal{J}(i,j)$ include $J_{1u}$, $J_{1v}$, $J_2$, and $J_3$, as illustrated in Fig.~\ref{Fig: Figure3}(c), where $J_{1u}$ and $J_{1v}$ denote the two inequivalent intra-dimer exchange interactions in the orthorhombic lattice~\cite{yadavObservationUnprecedentedFractional2025a}, $J_2$ and $J_3$ correspond to the inter-dimer and third-neighbor couplings, respectively. Besides the exchange parameters, we also considered the $B_2^0$ term in the CEF as a fitting parameter to account for the elongation distortion in this compound. Further calculation details of the neutron scattering cross section can be found in ~\ref{Appendix C}.

Using least squares fits, the coupling strengths and $B_2^0$ were refined to reproduce the excitation energies. Through the fits, we ensure that the resulting mean-field ground state at $T=0.05$~K is consistent with the experiments.  The resulting theoretical spectra for zero magnetic field are shown in Fig.~\ref{Fig: Figure5}(d), for which the fitted parameters are $J_{1u}=8.4(9)$, $J_{1v}=6.3(8)$, $J_{2}=2.4(5)$,  $J_{3}=1.2(2)$, and $B_2^0 = -2.5(8)$~$\mu$eV. Note that the exchange interactions in Ref.~\cite{yadavObservationUnprecedentedFractional2025a} are scaled by a factor of $|\hat{J}|\approx8$ compared to the values in our work. Besides the CEF excitation at \SI{1.57}{\milli\electronvolt}, the model accurately reproduces the three additional excitations that emerge from the ordered state: a low-energy magnon at \SI{0.37}{\milli\electronvolt} and two higher-energy modes at $1.83$ and \SI{1.98}{\milli\electronvolt}. Specifically, the energy of the \SI{0.37}{\milli\electronvolt} mode is set by the competition between the intra and inter-dimer couplings within the SSL geometry. Its weak dispersion and the robustness of its gap are consequences of the relatively strong Ising anisotropy. The high-energy modes at \SI{1.83}{\milli\electronvolt} and \SI{1.98}{\milli\electronvolt} arise from CEF transitions modified by the molecular field in the ordered state. In the paramagnetic state, the transition from the ground Kramers doublet to the first excited doublet occurs at \SI{1.57}{\milli\electronvolt}. Upon magnetic ordering, the first excited doublets on different sublattices split due to $J_{1u}\neq J_{1v}$, yielding the observed peaks at \SI{1.83} and \SI{1.98}{\milli\electronvolt}.
\begin{figure}
    \centering
    \includegraphics[width=\columnwidth]{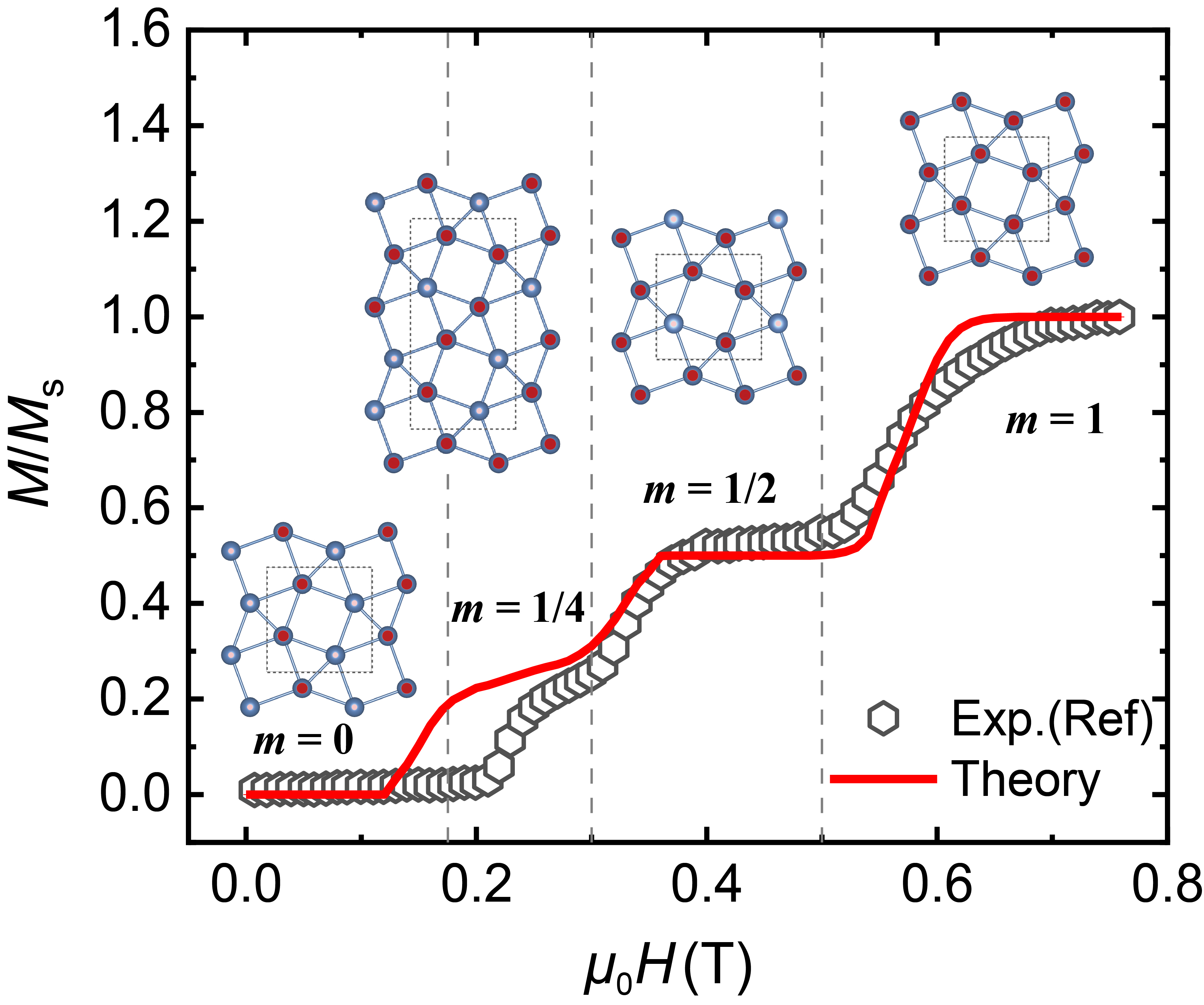}  
    \caption{Comparison between the experimental and calculated magnetization curves for $\ce{Er2Be2GeO7}$. The experimental data (hexagonal markers) were adapted from Fig.~11 of ref.~\cite{yadavObservationUnprecedentedFractional2025a}, which were measured at $T=0.3$~K in fields along the $c$-axis. Calculations (red line) were performed for the fitted spin Hamiltonian using the classical Monte Carlo simulations at $T = 0.675$~K, which is $\sim 0.36$ of the calculated $T_N=1.89$~K as that in experiments. The transitions into the fractional plateau phases and the saturation phase are indicated by vertical dashed lines. Insets are schematic illustrations of the spin configurations in each phase. For clarity, we simplified the spin states as spin-up (red circles) and spin-down (pink circles).}
    \label{Fig: Figure7} 
\end{figure} 

Using the fitted spin Hamiltonian, we calculate the magnetization curve via the classical Monte Carlo method. For this calculation, we work in the effective $S=\frac{1}{2}$ basis as in Ref.~\cite{yadavObservationUnprecedentedFractional2025a}, with the coupling strengths scaled from the fitted MF-RPA model (see ~\ref{Appendix D} for details). As shown in Fig.~\ref{Fig: Figure7}, the calculated magnetization curve reproduces the experimentally observed $m=\frac{1}{4}$ and $\frac{1}{2}$ fractional magnetization plateaus~\cite{yadavObservationUnprecedentedFractional2025a}. Like the previous CTMRG calculations~\cite{yadavObservationUnprecedentedFractional2025a}, our classical Monte Carlo simulations do not incorporate quantum fluctuations. Therefore, the agreement between the experimental and calculated magnetization curves suggests that the magnetization plateaus in \ce{Er2Be2GeO7} are mainly stabilized by Ising anisotropy as in ErB$_4$~\cite{huang_multi_2012, dublenych_ground_2012}, rather than by quantum fluctuations as in SrCu$_2$(BO$_3$)$_2$~\cite{dorier_theory_2008, abendschein_effective_2008, corboz_crystals_2014}.

Classical Monte Carlo simulations further determine the field-induced orders in the plateau phases. As shown by the insets in Fig.~\ref{Fig: Figure7}, we propose the corresponding spin configuration for the plateau phase based on the classical Ising model. At the $m = 1/2$ plateau, two out of the four antiferromagnetic dimers are excited and flipped into the ferromagnetic configuration, forming a unique up-up-up-down arrangement within the magnetic unit cell. Similar to the energy analysis in ref.~\cite{yadavObservationUnprecedentedFractional2025a}, our classical Monte Carlo simulations reveal a large number of degenerate states in the $m = 1/4$ plateau, which comprise all possible configurations with five up-pointing spins and three down-pointing spins. 

\section{Conclusion}
\label{sec:Conclusion}

The magnetic structure and zero-field critical behavior of \ce{Er2Be2GeO7} establish that its magnetic ground state is governed by strong Ising anisotropy. Within the ordered phase, the excitation spectrum exhibits a finite energy gap and weak momentum dispersion, well reproduced by our MF-RPA calculations. Importantly, reproducing the inelastic neutron scattering spectra requires inequivalent nearest-neighbor interactions ($J_{1u} \neq J_{1v}$), which provides further evidence for orthorhombic distortion in \ce{Er2Be2GeO7}. The field-induced fractional magnetization plateaus are reproduced within classical Monte Carlo simulations using the exchange parameters determined from the INS spectra. Taken together, these results establish a unified description of the spin correlations in  \ce{Er2Be2GeO7}, in which the spin dynamics are quantitatively captured by the MF-RPA framework, while the static magnetic ordering and field-induced plateau phases are governed by dominant Ising anisotropy.

The emergence of classical behavior identified in \ce{Er2Be2GeO7} is consistent with a recurring theme in quantum magnets, where prominent semi-classical behavior can emerge within the ordered regime~\cite{ran_spin_2017, samarakoon_comparative_2017, yao_excitations_2022}. Although the SSL geometry is often associated with strong quantum fluctuations, the strong single-ion anisotropy of the \ce{Er^{3+}} moments constrains the low-energy degrees of freedom predominantly along their Ising axes. Quantum CEF matrix elements remain essential for describing the INS response, whereas the static magnetization process, including the \(1/4\) and \(1/2\) plateaus, is well captured by an anisotropic classical Ising model.

\appendix
\section{Comparison of tetragonal and orthorhombic structural models}
\label{Appendix A}
\begin{figure}[h]
    \centering
        \includegraphics[width=\columnwidth]{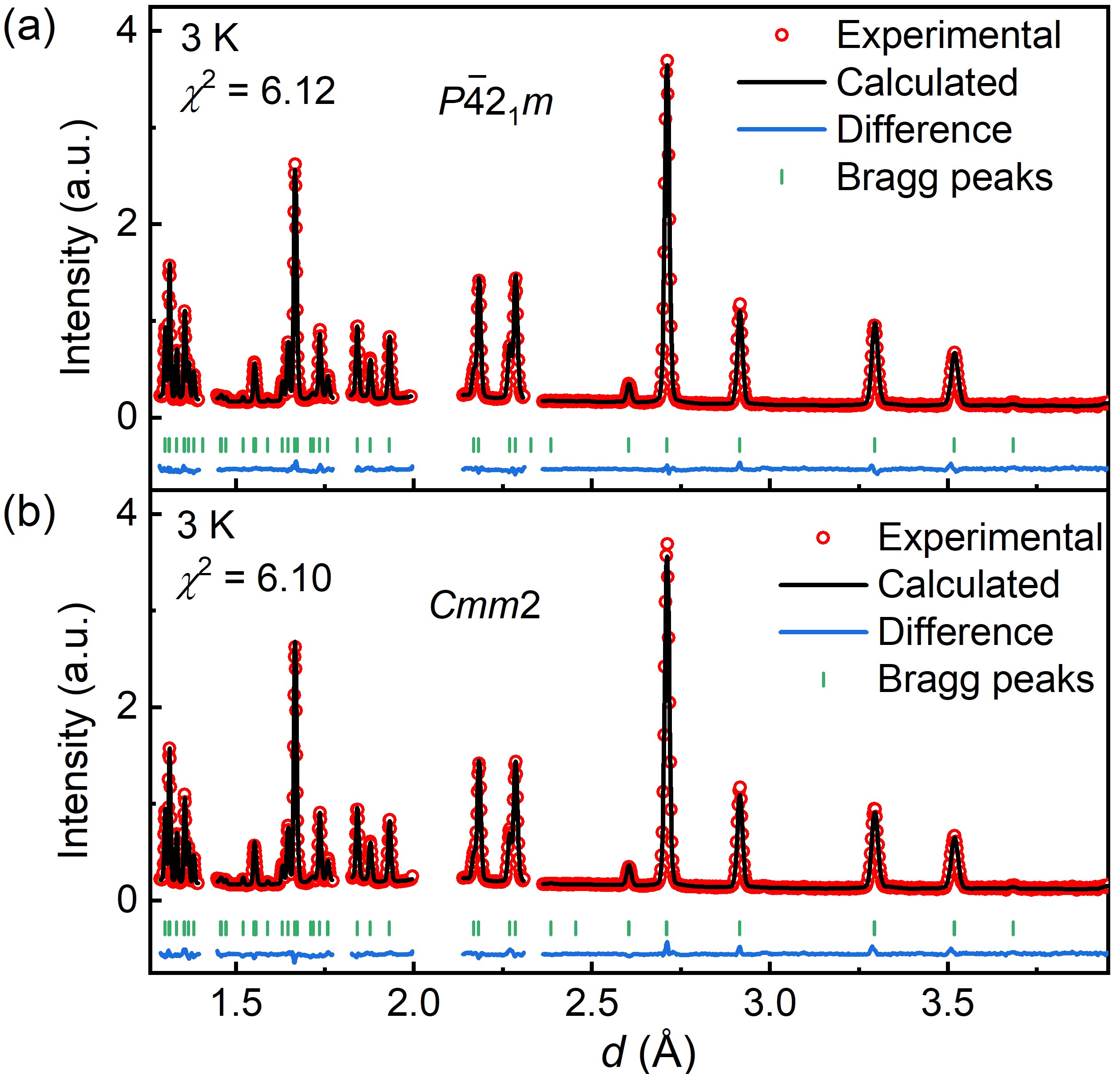}
        \caption{Rietveld refinements of the TOF-NPD pattern at 3 K of the SSL compound \ce{Er2Be2GeO7}, using the $P\overline{4}2_1m$ (a) and $Cmm2$ (b) structural models. Red circles, black line and blue line are the experimental, calculated intensities and difference, respectively.  Green ticks represent the positions of the Bragg reflections.}
        \label{App:1}
\end{figure}

To address the low-temperature structural properties and the orthorhombic distortion ($Cmm2$ space group) reported in single-crystal studies~\cite{yadavObservationUnprecedentedFractional2025a}, we performed a comparative Rietveld analysis on the NPD data collected at \SI{3}{\kelvin}. In principle, the symmetry lowering from $P\overline{4}2_1m$ to $Cmm2$ ($a \neq b$) lifts the equivalence of the orthogonal dimers and is expected to produce a splitting of certain Bragg reflections. However, the nuclear Bragg peaks splitting was not observed in NPD patterns at any temperatures, as well as the additional peaks of \{1 0 0\}~\cite{yadavObservationUnprecedentedFractional2025a}. With this consideration, the Rietveld refinement using the lower-symmetry $Cmm2$ model yields inequal lattice parameters of $a = 10.4066(09)$~\AA\ and $b = 10.4029(09)$~\AA, corresponding to the relative in-plane orthorhombic distortion of $\Delta a / a \approx 3.5 \times 10^{-4}$. This value is substantially smaller than the instrument $d$-spacing resolution of the General Purpose Powder Diffractometer (GPPD) ($\Delta d/d \sim 2.0 \times 10^{-3}$)~\cite{gppd}, such that the associated orthorhombic peak splitting cannot be resolved in the present powder diffraction data. As shown in Fig.~\ref{App:1}, refinements using both the high-symmetry $P\overline{4}2_1m$ model and the lower-symmetry $Cmm2$ model describe the experimental NPD data equally well, yielding nearly identical profiles and refinement quality indicators. Therefore, the tetragonal $P\overline{4}2_1m$ space group serves as the global approximation of our crystallographic analysis in a simple way.

\section{Inelastic spectrum: damped harmonic oscillator (DHO) fits}
\label{Appendix B}
To quantify the evolution of these excitations, the momentum-integrated spectra were analyzed using a Gaussian function for the elastic scattering contribution together with damped harmonic oscillator (DHO) functions for the inelastic magnetic excitations. The fitting function is given by:
\begin{equation}
\begin{aligned}
I(Q, E) = {} & A \cdot \mathrm{IRF}(Q, E) \\
& + \frac{1}{1 - \exp\left(-\frac{E}{k_{\mathrm{B}} T}\right)} \sum_i \frac{C_i E \Gamma_i}{(E^2 - E_i^2)^2 + (E \Gamma_i)^2}
\end{aligned}
\end{equation}
where the first term represents the elastic Gaussian contribution, and the second term accounts for multiple inelastic magnetic components. Here, $E_\mathrm{i}$ denotes the peak energy of the $i$-th excitation, $\Gamma_\mathrm{i}$ its full width at half maximum (FWHM), and $C_\mathrm{i}$ an intensity prefactor, while the Bose population factor ensures detailed balance.

\begin{figure}[htbp]
    \centering
    \includegraphics[width=\columnwidth]{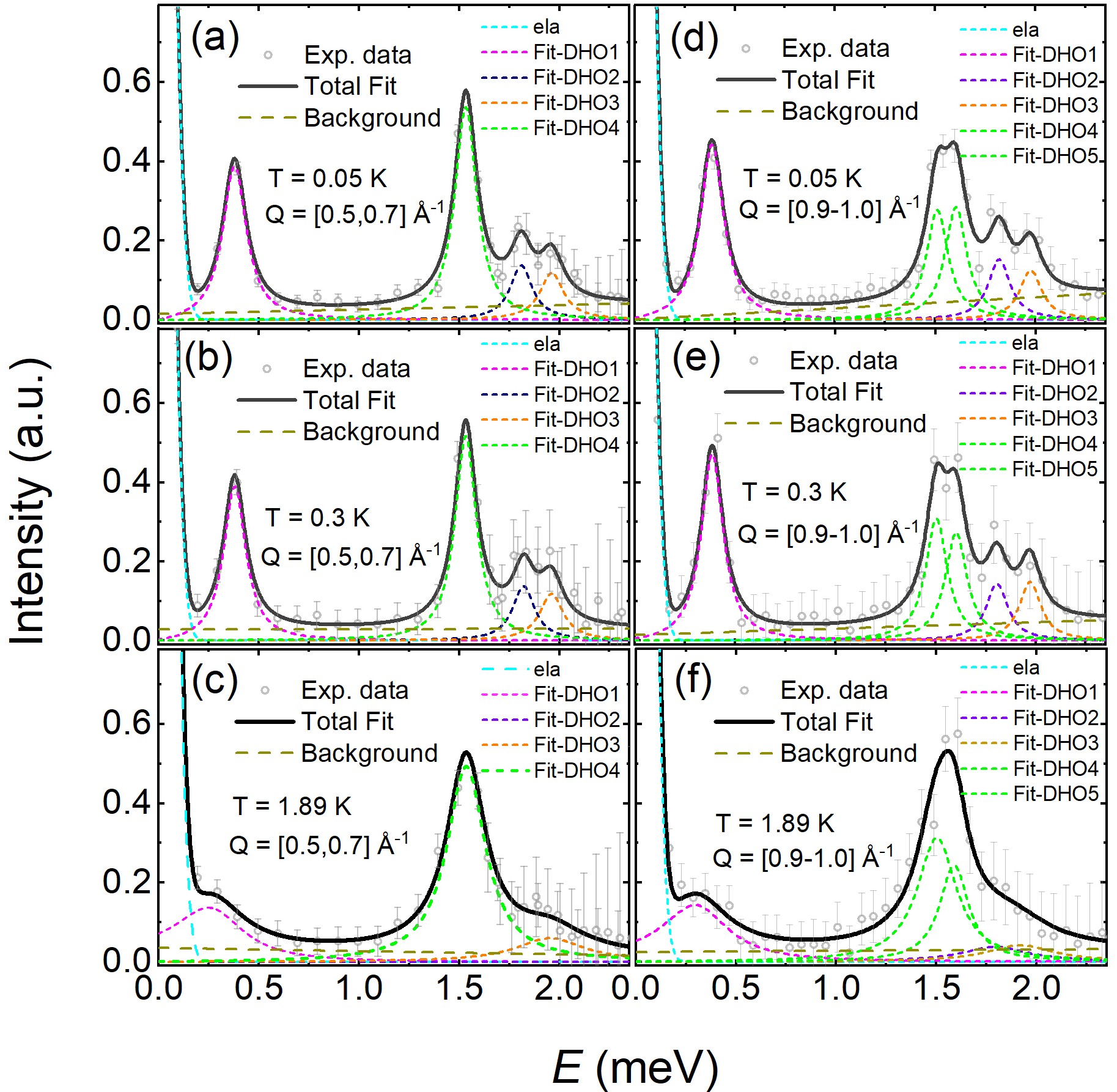}  
    \caption{INS spectra acquired at temperatures of 0.05, 0.3, and 1.89~K for momentum transfers of $Q = 0.5-0.7~\text{\AA}^{-1}$ (a-c) and $Q = 0.9-1.0~\text{\AA}^{-1}$ (d-f).  Closed circles represent the experimental data. The solid lines show the corresponding total fits obtained using the DHO model. Individual excitation peaks are described by dashed curves in different colors. The fits comprehensively capture the evolution of multiple excitation peaks as a function of  temperature and momentum transfer.} 
    \label{App:2} 
\end{figure}

Representative fits are displayed in Fig.~\ref{App:2} for spectra integrated over two momentum-transfer ranges at selected temperatures. The experimental data are well described by the DHO model across all temperatures and $Q$ ranges. The consistent quality of these fits validates the extracted temperature and field dependencies of the peak energies ($E_i$) and linewidths ($\Gamma_i$) reported in the main text. The distinct evolution of these parameters for the \SI{0.37}{\milli\electronvolt} mode compared to the higher-energy doublet forms the quantitative basis for our interpretation of their different physical origins (collective magnon and molecular field split CEF excitations). An additional nuance is observed for the $\SI{1.57}{\milli\electronvolt}$ excitation at low temperature. While it appears as a single, $Q$-independent peak in the paramagnetic state, its lineshape at $T < T_N$ acquires a subtle $Q$ dependence. DHO analysis reveals that at higher $Q$ ($Q = 0.9-1.0~\text{\AA}^{-1}$), the spectrum is best described by two closely spaced components, unlike the single component sufficient at lower $Q$. This reflects a redistribution of spectral weight among several closely spaced transitions. In the ordered state, the four inequivalent magnetic sublattices experience slightly different molecular fields, causing small shifts of the CEF levels on each sublattice. The measured, momentum-dependent lineshape thus arises from the differential visibility of these site-split components, providing further microscopic evidence for the ordered magnetic structure.

\section{MF-RPA method}
\label{Appendix C}
The Hamiltonian of the system is expressed as:
\begin{equation}
    \mathcal{H} = \sum_i \mathcal{H}_{\mathrm{SI}}^i + \mathcal{H}_{\mathrm{int}},\label{eq:Htot}
\end{equation}
where $\mathcal{H}_{\mathrm{SI}}^i$ is the single-ion Hamiltonian at site $i$ and $\mathcal{H}_{\mathrm{int}}$ describes the inter-ion exchange interactions. The single-ion Hamiltonian for site $i$ is determined by the CEF, $\mathcal{H}_{\mathrm{SI}}^i = \mathcal{H}_{\mathrm{CEF}}^i=\sum_{k=2,4,6} \sum_{q=-k}^{k} B_k^q O_k^q$. For each of the four sublattices ($s$) in a unit cell, a local coordinate frame was defined with the $\tilde{z}_s$ axis being parallel with the  ordered moment while keeping a positive projection along the crystallographic $z$ axis. For each sublattice, the local angular momentum operators, $\tilde{\bm{J}^{s}}$, are linked to the corresponding global angular momentum operators, $\bm{J}^{s}$, through a rotation matrix $R^s$, such that $R^{s}\tilde{\bm{J}^{s}}=\bm{J}^{s}$. From $R^{s}$, a transform operator, $\mathcal{U}^s$, can be defined, which rotates the CEF Hamiltonian to the local frame,
\begin{equation}
    \tilde{\mathcal{H}}^s_{\mathrm{CEF}} = \mathcal{U}^s \mathcal{H}^s_{\mathrm{CEF}}\mathcal{U}^{s\dagger}.
\end{equation}
The exchange interactions among the \ce{Er^{3+}} ions are described by a general bilinear anisotropic Hamiltonian in the local frame:
\begin{equation}
    \mathcal{H}_{\mathrm{int}} = \frac{1}{2}\sum_{i \neq j}\sum_{\alpha\beta}\tilde{\mathcal{J}}^{ss'}_{\alpha\beta}(\Delta \mathbf{R}_{ij})\tilde{J}_\alpha^i\tilde{J}_\beta^{j}.
\end{equation}
Here,
$\tilde{\mathcal{J}}_{\alpha\beta}$ is the exchange tensor defined in the local frame between sites $i$ and $j$ separated by the vector $\Delta\mathbf{R}_{ij}=\mathbf{R}_{i}-\mathbf{R}_{j}$. Under the MF approximation, the Hamiltonian for each sublattice is then solved self-consistently in the rotating local frame formalism~\cite{PhysRevB.106.054431}:
\begin{align}
    &\tilde{\mathcal{H}}_\mathrm{MF}^s = \tilde{\mathcal{H}}_{\mathrm{CEF}}^s - \sum_{\alpha}\tilde{H}_\alpha^s \tilde{J}_\alpha^s, \\
    &\tilde{H}_\alpha^s = \sum_{s',\beta} \tilde{\mathcal{J}}_{\alpha\beta}^{ss'} \langle\tilde{J}_{\beta}^{s'}\rangle.
\end{align}
Given the strong Ising-like anisotropy observed in experiments, we restrict the exchange tensors to the Ising form, $\tilde{\mathcal{J}}^{ss'}_{\alpha\beta} = \tilde{\mathcal{J}}_{\alpha\beta}\delta^{s}_{\alpha \tilde{z}}\delta^{s'}_{\beta \tilde{z}}$. Following the analysis in Ref.~\cite{yadavObservationUnprecedentedFractional2025a}, we consider the following couplings as indicated in Fig.~\ref{Fig: Figure3}(c) of the main text: two disparate intra-dimer interactions $J_{1u}$ and $J_{1v}$ due to the lattice distortion, together with the inter-dimer interactions $J_2$ and $J_3$ at further neighbors. 

From the MF ground state, we can calculate the single-ion dynamical susceptibility for each sublattice following the linear response theory:
\begin{equation}
    \small\tilde{\chi}_{\alpha\beta}^{0,s}(\omega)=\sum_{m,n}\frac{\langle m_s|\tilde{J}_\alpha^s-\langle\tilde{J}_\alpha^s\rangle|n_s\rangle\langle n_s|\tilde{J}_\beta^s-\langle\tilde{J}_\beta^s\rangle|m_s\rangle}{E_{n_s}-E_{m_s}-\hbar\omega-i\epsilon}(p_{m_s}-p_{n_s}).
\end{equation}
Here, $|m_s\rangle$ and $E_{m_s}$ are the eigenstates and eigenvalues of $\tilde{\mathcal{H}}_\mathrm{MF}^s$, and $p_{m_s}$ is the corresponding Boltzmann population factor.

Under the random phase approximation~\cite{Jensen:RareEarth,Rotter_2012}, the total susceptibility tensor in the local frame, $\tilde{\chi}(\mathbf{Q},\omega)$, is given by the matrix equation:
\begin{equation}
    \tilde{\chi}(\mathbf{Q},\omega) = \left[ (\tilde{\chi}^0(\omega))^{-1} - \tilde{\mathcal{J}}(\mathbf{Q}) \right]^{-1},
    \label{eq:RPA}
\end{equation}
where $\tilde{\chi}^0(\omega)$ is a block-diagonal matrix of the single-ion susceptibilities, $\tilde{\chi}^0(\omega) = \bigoplus_{s=1}^4 \tilde{\chi}_{\alpha\beta}^{0,s}$, and $\tilde{\mathcal{J}}(\mathbf{Q})$ is the Fourier transform of the exchange tensor in the rotating frame. For comparison with the experimental results, the total susceptibility in the local frame can be rotated back to the global frame  using the transformation $\chi(\mathbf{Q},\omega) = \mathcal{R} \cdot \tilde{\chi}(\mathbf{Q},\omega) \cdot \mathcal{R}^T$, where $\mathcal{R}=\bigoplus_{s=1}^4 R^s$~\cite{PhysRevB.106.054431}.

The spectral function, $S(\mathbf{Q},\omega)$, which is directly probed in the INS experiments, is calculated from the total susceptibility in the global frame via the fluctuation-dissipation theorem
\begin{equation}
    \begin{aligned}
    S(\mathbf{Q},\omega)\propto \frac{F^2(Q)}{1-e^{-\hbar\omega/k_\mathrm{B} T}} \sum_{\alpha,\beta}(\delta_{\alpha\beta}-\hat{Q}_{\alpha}\hat{Q}_{\beta}) 
     \mathrm{Im}\left[\chi_{\alpha\beta}(\mathbf{Q},\omega)\right],
    \end{aligned}
\end{equation}
where $F(Q)$ is the magnetic form factor of the \ce{Er^{3+}} ion. For comparisons with the experiments, we performed a powder average of the calculated $S(\mathbf{Q},\omega)$:
\begin{equation}
    S(Q,E) = \int S(\mathbf{Q},\omega) \, d\Omega_{\mathbf{Q}}/4\pi.
\end{equation}
In our calculations, the powder average was obtained by sampling 200 random points at each $Q$ position. The calculated spectra were convoluted with a constant energy resolution of 0.1~meV.

The emergence of the 1/4 magnetization plateau necessitates a spatial anisotropy in the intra-dimer couplings ($J_{1u} \neq J_{1v}$). We find that this conclusion is highly consistent with our INS spectra. An isotropic model ($J_{1u} = J_{1v}$) yields a single band above 1.6~meV. When the spatially anisotropic condition ($J_{1u} \neq J_{1v}$) is introduced, the broken sublattice equivalence splits the upper bands. The parameters from a recent study~\cite{yadavObservationUnprecedentedFractional2025a}, for instance, feature a larger anisotropy between $J_{1u}$ and $J_{1v}$, so that the gap between split bands is not consistent with the experiment. Furthermore, the inter-dimer coupling, $J_2$, dramatically affects the momentum-dependent dispersion of the upper three bands and must therefore remain relatively small to avoid introducing excessive dispersion. Finally, the third-neighbor interaction, $J_3$, serves as a crucial tuning parameter that dictates the overall energy gap of all the bands while also refining the dispersion of the upper three bands.

\section{Monte Carlo simulation}
\label{Appendix D}
Classical MC simulations were performed on a $20 \times 20 \times 1$ supercell with periodic boundary conditions using the SpinMC code that implements the single spin flip Metropolis algorithm~\cite{buessen_spinmc_2020}. Based on the Ising-type anisotropy derived from the fitted crystal electric field, classical Ising spins were utilized. Magnetization was averaged over $1\times10^5$ of measurement sweeps after $5\times10^5$ thermalization sweeps, where each sweep represents 1600 attempted spin flips at randomly selected sites. The parallel tempering algorithm was utilized to facilitate thermal equilibrium, which was performed simultaneously over 80 replicas on 80 cores with a geometric series of temperatures between 0.1 and 4~K. After every 10 Monte Carlo sweeps, a replica exchange was attempted, and the successful exchange rates are above $\sim25\%$ for all the neighboring replicas. The magnetization curve at $T=0.675$~K was shown in the main text.
\begin{equation}
\mathcal{H}
=
\sum_{\langle ij\rangle=1u,1v,2,3}
J_{ij}|J|^2S_i^zS_j^z-g\mu_{\mathrm B}H |J|\sum_i S_i^z,
\label{eq:model}
\end{equation}
Here, \(J_{ij}\) denotes the exchange parameter defined for the full angular-momentum operators used in the MF-RPA calculation, whereas \(S_i^z=\pm1\) is a normalized classical Ising variable. When compared with a unit-spin convention, the exchange scale inferred from the critical fields differs by one net factor of \(\lvert{J}\rvert\approx8\). The apparent difference between the reported exchange parameters in Ref.~\cite{yadavObservationUnprecedentedFractional2025a} results from the spin-normalization convention rather than from different underlying interactions.

\section*{CRediT authorship contribution statement}
Donger Cheng: Data curation, Formal analysis, Methodology, Validation, Visualization, Writing -- original draft, Writing -- review \& editing. Otkur Omar: Data curation, Formal analysis, Methodology, Validation, Visualization, Writing -- original draft, Writing -- review \& editing. Wenbin Wei: Investigation. Jan Peter Embs: Data curation, Investigation, Visualization. Xiao Wang: Funding acquisition, Investigation, Supervision, Writing -- original draft, Writing -- review \& editing. Cheng Su: Investigation. Hui Cheng: Investigation. Zhengdong Fu: Data curation, Investigation, Resources. Xinzhi Liu: Investigation, Supervision. Wentao Jin: Investigation, Resources. Ze Hu: Formal analysis, Visualization. Wei Luo: Formal analysis. Liming Wang: Investigation, Resources. Jiazheng Hao: Formal analysis, Visualization. Feiran Shen: Formal analysis. Lunhua He: Data curation, Investigation. Shang Gao: Conceptualization, Data curation, Formal analysis, Funding acquisition, Investigation, Methodology, Resources, Supervision, Validation, Visualization, Writing -- original draft, Writing -- review \& editing. Erxi Feng: Conceptualization, Data curation, Formal analysis, Funding acquisition, Investigation, Methodology, Project administration, Resources, Supervision, Validation, Visualization, Writing -- original draft, Writing -- review \& editing.
\section*{Declaration of competing interest}

The authors declare that they have no known competing financial
interests or personal relationships that could have appeared to
influence the work reported in this paper.

\section*{Acknowledgements}

We acknowledge helpful discussions with Huike Jin. This work was supported by the National Natural Science Foundation of China (NSFC) under Grant Nos. 12375298, 12425512, and 12505350, and by the Large Scientific Facility Open Subject of Songshan Lake, Dongguan, Guangdong, KFKT2022B05. Work at the University of Science and Technology of China was supported by the National Key R\&D Program of China under Grant No. 2024YFA1613100 and the NSFC under Grant No. 12374152.


\section*{Data availability}

Data will be made available on request.

\bibliographystyle{elsarticle-num-doiurl}
\bibliography{references}

@PREAMBLE{
 "\providecommand{\noopsort}[1]{}" 
 # "\providecommand{\singleletter}[1]{#1}%" 
}

@article{balentsSpinLiquidsFrustrated2010,
  title = {Spin Liquids in Frustrated Magnets},
  author = {Balents, Leon},
  year = {2010},
  month = mar,
  journal = {Nature},
  volume = {464},
  number = {7286},
  pages = {199--208},
  issn = {0028-0836, 1476-4687},
  doi = {10.1038/nature08917},
  urldate = {2025-07-28},
  langid = {english},
}

@article{broholmQuantumSpinLiquids2020a,
  title = {Quantum Spin Liquids},
  author = {Broholm, C. and Cava, R. J. and Kivelson, S. A. and Nocera, D. G. and Norman, M. R. and Senthil, T.},
  year = {2020},
  month = jan,
  journal = {Science},
  volume = {367},
  number = {6475},
  issn = {0036-8075, 1095-9203},
  doi = {10.1126/science.aay0668},
  pages = {eaay0668},
  urldate = {2025-08-25},
  langid = {english},
}

@article{sriramshastryExactGroundState1981a,
  title = {{{Exact Ground State}} of a {{Quantum Mechanical Antiferromagnet}}},
  author = {Shastry, B. S. and Sutherland, B.},
  year = {1981},
  month = aug,
  journal = {Physica B+C},
  volume = {108},
  number = {1-3},
  pages = {1069--1070},
  issn = {03784363},
  doi = {10.1016/0378-4363(81)90838-X},
  urldate = {2024-10-17},
  copyright = {https://www.elsevier.com/tdm/userlicense/1.0/},
  langid = {english},
}

@article{knetterDispersionSymmetryBound2000,
  title = {Dispersion and {{Symmetry}} of {{Bound States}} in the {{Shastry-Sutherland Model}}},
  author = {Knetter, Christian and Bühler, Alexander and Müller-Hartmann, Erwin and Uhrig, Götz S.},
  year = {2000},
  journal = {Phys. Rev. Lett.},
  volume = {85},
  number = {18},
  pages = {3958--3961},
  publisher = {American Physical Society},
  doi = {10.1103/PhysRevLett.85.3958},
  urldate = {2024-06-24}
}

@article{kageyamaCrystalGrowthTwodimensional1999,
  title = {Crystal {{Growth}} of the {{Two-Dimensional Spin Gap System}} \ce{SrCu2(BO3)2}},
  author = {Kageyama, H and Onizuka, K and Yamauchi, T and Ueda, Y},
  year = {1999},
  month = oct,
  journal = {J. Cryst. Growth},
  volume = {206},
  number = {1-2},
  pages = {65--67},
  issn = {00220248},
  doi = {10.1016/S0022-0248(99)00313-9},
  urldate = {2024-10-17},
  copyright = {https://www.elsevier.com/tdm/userlicense/1.0/},
  langid = {english}
}

@article{mullerhartmannExactDemonstrationMagnetization2000,
  title = {{{Exact Demonstration}} of {{Magnetization Plateaus}} and {{{First-Order Dimer-Neel Phase Transitions}}} in a {Modified Shastry-Sutherland Model} for \ce{SrCu2(BO3)2}},
  author = {{M{\"u}ller-Hartmann}, Erwin and Singh, Rajiv R. P. and Knetter, Christian and Uhrig, G{\"o}tz S.},
  year = {2000},
  month = feb,
  journal = {Phys. Rev. Lett.},
  volume = {84},
  number = {8},
  pages = {1808--1811},
  issn = {0031-9007, 1079-7114},
  doi = {10.1103/PhysRevLett.84.1808},
  urldate = {2024-10-17},
  copyright = {http://link.aps.org/licenses/aps-default-license},
  langid = {english},
}

@article{kodamaMagneticSuperstructureTwoDimensional2002a,
  title = {{{Magnetic Superstructure}} in the {{Two-Dimensional Quantum Antiferromagnet}} \ce{SrCu2(BO3)2}},
  author = {Kodama, K. and Takigawa, M. and Horvati{\'c}, M. and Berthier, C. and Kageyama, H. and Ueda, Y. and Miyahara, S. and Becca, F. and Mila, F.},
  year = {2002},
  month = oct,
  journal = {Science},
  volume = {298},
  number = {5592},
  pages = {395--399},
  issn = {0036-8075, 1095-9203},
  doi = {10.1126/science.1075045},
  urldate = {2025-08-25},
  langid = {english},
}

@article{kageyamaDirectEvidenceLocalized2000,
  title = {Direct {{Evidence}} for the {{Localized Single-Triplet Excitations}} and the {{Dispersive Multitriplet Excitations}} in  \ce{SrCu2(BO3)2}},
  author = {Kageyama, H. and Nishi, M. and Aso, N. and Onizuka, K. and Yosihama, T. and Nukui, K. and Kodama, K. and Kakurai, K. and Ueda, Y.},
  year = {2000},
  date = {2000-06-19},
  journal = {Phys. Rev. Lett.},
  shortjournal = {Phys. Rev. Lett.},
  volume = {84},
  number = {25},
  pages = {5876--5879},
  publisher = {American Physical Society},
  doi = {10.1103/PhysRevLett.84.5876},
  urldate = {2024-10-17}
}

@article{kageyamaExactDimerGround1999c,
  title = {{{Exact Dimer Ground State}} and {{Quantized Magnetization Plateaus}} in the {{Two-Dimensional Spin System}} \ce{SrCu2(BO3)2}},
  author = {Kageyama, H. and Yoshimura, K. and Stern, R. and Mushnikov, N. V. and Onizuka, K. and Kato, M. and Kosuge, K. and Slichter, C. P. and Goto, T. and Ueda, Y.},
  year = {1999},
  date = {1999-04-12},
  journal = {Phys. Rev. Lett.},
  volume = {82},
  number = {15},
  pages = {3168--3171},
  publisher = {American Physical Society},
  doi = {10.1103/PhysRevLett.82.3168},
  urldate = {2024-10-17},
  langid = {english}
}

@article{lemmensCollectiveSingletExcitations2000,
  title = {{{Collective Singlet Excitations}} and {{Evolution}} of {{Raman Spectral Weights}} in the {{2D Spin Dimer Compound}} \ce{SrCu2(BO3)2}},
  author = {Lemmens, P. and Grove, M. and Fischer, M. and G{\"u}ntherodt, G. and Kotov, Valeri N. and Kageyama, H. and Onizuka, K. and Ueda, Y.},
  year = {2000},
  month = sep,
  journal = {Phys. Rev. Lett.},
  volume = {85},
  number = {12},
  pages = {2605--2608},
  issn = {0031-9007, 1079-7114},
  doi = {10.1103/PhysRevLett.85.2605},
  urldate = {2024-10-17},
  copyright = {http://link.aps.org/licenses/aps-default-license},
  langid = {english},
}

@article{nomuraUnveilingNewQuantum2023,
  title = {{{Unveiling New Quantum Phases}} in the {{Shastry-Sutherland Compound}} \ce{SrCu2(BO3)2} up to the {{Saturation Magnetic Field}}},
  author = {Nomura, T. and Corboz, P. and Miyata, A. and Zherlitsyn, S. and Ishii, Y. and Kohama, Y. and Matsuda, Y. H. and Ikeda, A. and Zhong, C. and Kageyama, H. and Mila, F.},
  year = {2023},
  month = jun,
  journal = {Nat. Commun.},
  volume = {14},
  number = {1},
  pages = {3769},
  publisher = {Nature Publishing Group},
  issn = {2041-1723},
  doi = {10.1038/s41467-023-39502-5},
  urldate = {2024-05-24},
  copyright = {2023 The Author(s)},
  langid = {english},
}

@article{zayed4spinPlaquetteSinglet2017,
  title = {{{4-Spin Plaquette Singlet State}} in the {{Shastry--Sutherland Compound}} \ce{SrCu2(BO3)2}},
  author = {Zayed, M. E. and R{\"u}egg, Ch and Larrea J., J. and L{\"a}uchli, A. M. and Panagopoulos, C. and Saxena, S. S. and Ellerby, M. and McMorrow, D. F. and Str{\"a}ssle, Th and Klotz, S. and Hamel, G. and Sadykov, R. A. and Pomjakushin, V. and Boehm, M. and {Jim{\'e}nez--Ruiz}, M. and Schneidewind, A. and Pomjakushina, E. and Stingaciu, M. and Conder, K. and R{\o}nnow, H. M.},
  year = {2017},
  month = oct,
  journal = {Nat. Phys.},
  volume = {13},
  number = {10},
  pages = {962--966},
  publisher = {Nature Publishing Group},
  issn = {1745-2481},
  doi = {10.1038/nphys4190},
  urldate = {2024-06-24},
  copyright = {2017 Springer Nature Limited},
  langid = {english}
}

@article{kim$textYb_2textPt_2textPb$MagneticFrustration2008,
  title = {\ce{Yb2Pt2Pb}: {{Magnetic Frustration}} in the {{Shastry-Sutherland Lattice}}},
  shorttitle = {\$\{{\textbackslash}text\{\vphantom{\}\}}{{Yb}}\vphantom\{\}\vphantom\{\}\_\{2\}\{{\textbackslash}text\{\vphantom{\}\}}{{Pt}}\vphantom\{\}\vphantom\{\}\_\{2\}{\textbackslash}text\{\vphantom\}{{Pb}}\vphantom\{\}\$},
  author = {Kim, M. S. and Bennett, M. C. and Aronson, M. C.},
  year = {2008},
  month = apr,
  journal = {Phys. Rev. B},
  volume = {77},
  number = {14},
  pages = {144425},
  publisher = {American Physical Society},
  doi = {10.1103/PhysRevB.77.144425},
  urldate = {2024-06-25},
  langid = {english}
}

@article{matsumuraNonCollinearMagneticStructure2007a,
  title = {Non-{{Collinear Magnetic Structure}} of \ce{TbB4}},
  author = {Matsumura, Takeshi and Okuyama, Daisuke and Murakami, Youichi},
  year = {2007},
  month = jan,
  journal = {J. Phys. Soc. Jpn.},
  volume = {76},
  number = {1},
  pages = {015001},
  publisher = {The Physical Society of Japan},
  issn = {0031-9015},
  doi = {10.1143/JPSJ.76.015001},
  urldate = {2025-03-31},
  langid = {english},
}

@article{metokiMagneticQuadrupoleCoupling2025,
  title = {Magnetic and {{Quadrupole Coupling Emerging}} in \ce{NdB4} with {{Geometrically Frustrated Orthogonal Antiferromagnetic Dimer}}},
  author = {Metoki, Naoto and Yamauchi, Hiroki and Hagihala, Masato and Watanuki, Ryuta and Kawamura, Seiko and Kofu, Maiko and Nakajima, Kenji and Matsuda, Masaaki},
  year = {2025},
  month = mar,
  journal = {Phys. Rev. B},
  volume = {111},
  number = {10},
  pages = {104424},
  publisher = {American Physical Society},
  doi = {10.1103/PhysRevB.111.104424},
  urldate = {2025-03-23},
}

@article{shiDiscoveryQuantumPhases2022,
  title = {Discovery of {{Quantum Phases}} in the {{Shastry-Sutherland Compound}} \ce{SrCu2(BO3)2} under {{Extreme Conditions}} of {{Field}} and {{Pressure}}},
  author = {Shi, Zhenzhong and Dissanayake, Sachith and Corboz, Philippe and Steinhardt, William and Graf, David and Silevitch, D. M. and Dabkowska, Hanna A. and Rosenbaum, T. F. and Mila, Frédéric and Haravifard, Sara},
  year = {2022},
  date = {2022-04-28},
  journal = {Nat. Commun.},
  volume = {13},
  number = {1},
  pages = {2301},
  publisher = {Nature Publishing Group},
  issn = {2041-1723},
  doi = {10.1038/s41467-022-30036-w},
  urldate = {2024-06-24}
}

@article{ashtarStructureMagneticProperties2021b,
  title = {Structure and {{Magnetic Properties}} of {{Melilite-Type Compounds}} \ce{Re2Be2GeO7} {(RE = Pr, Nd, Gd--Yb)} with {{Rare-Earth Ions}} on {{Shastry--Sutherland Lattice}}},
  author = {Ashtar, Malik and Bai, Yuming and Xu, Longmeng and Wan, Zongtang and Wei, Zijun and Liu, Yong and Marwat, Mohsin Ali and Tian, Zhaoming},
  year = {2021},
  volume = {60},
  issue = {6},
  pages = {3626-3634},
  journal = {Inorg. Chem.},
  doi = {10.1021/acs.inorgchem.0c03131},
  langid = {english},
}

@article{liuDistinctMagneticGround2024a,
  title = {Distinct {{Magnetic Ground States}} in {{Shastry-Sutherland Lattice Materials}}: \ce{Pr2Be2GeO7} versus \ce{Nd2Be2GeO7}},
  shorttitle = {Distinct Magnetic Ground States in {{Shastry-Sutherland}} Lattice Materials},
  author = {Liu, Andi and Song, Fangyuan and Cao, Yaotao and Ge, Han and Bu, Huanpeng and Zhou, Jin and Qin, Yuqi and Zeng, Qingyuan and Li, Jingxin and Ling, Langsheng and Tong, Wei and Sheng, Jieming and Yang, Ming and Wu, Liusuo and Guo, Hanjie and Tian, Zhaoming},
  year = {2024},
  month = may,
  journal = {Phys. Rev. B},
  volume = {109},
  number = {18},
  pages = {184413},
  publisher = {American Physical Society},
  doi = {10.1103/PhysRevB.109.184413},
  urldate = {2024-05-24},
  langid = {english},
}

@article{liuLargeMagnetocaloricEffect2024,
  title = {Large {{Magnetocaloric Effect}} in the {{Shastry-Sutherland Lattice Compound}} \ce{Yb2Be2GeO7} with {{Spin-Disordered Ground State}}},
  author = {Liu, Andi and Zhou, Jin and Wang, Lei and Cao, Yantao and Song, Fangyuan and Han, Yuyan and Li, Jingxin and Tong, Wei and Xia, Zhengcai and Ouyang, Zhongwen and Zhao, Jinkui and Guo, Hanjie and Tian, Zhaoming},
  year = {2024},
  month = oct,
  journal = {Phys. Rev. B},
  volume = {110},
  number = {14},
  pages = {144445},
  issn = {2469-9950, 2469-9969},
  doi = {10.1103/PhysRevB.110.144445},
  urldate = {2024-11-18},
  langid = {english},
}

@article{mccuskerRietveldRefinementGuidelines1999,
  title = {Rietveld Refinement Guidelines},
  author = {McCusker, L. B. and Von Dreele, R. B. and Cox, D. E. and Lou{\"e}r, D. and Scardi, P.},
  year = {1999},
  month = feb,
  journal = {J. Appl. Crystallogr.},
  volume = {32},
  number = {1},
  pages = {36--50},
  issn = {0021-8898},
  doi = {10.1107/S0021889898009856},
  urldate = {2025-07-28},
}

@article{azuahDAVEComprehensiveSoftware2009,
  title = {{{DAVE}}: {{A Comprehensive Software Suite}} for the {{Reduction}}, {{Visualization}}, and {{Analysis}} of {{Low Energy Neutron Spectroscopic Data}}},
  shorttitle = {{{DAVE}}},
  author = {Azuah, Richard Tumanjong and Kneller, Larry R. and Qiu, Yiming and {Tregenna-Piggott}, Philip L. W. and Brown, Craig M. and Copley, John R. D. and Dimeo, Robert M.},
  year = {2009},
  month = nov,
  journal = {J. Res. Natl. Inst. Stand. Technol.},
  volume = {114},
  number = {6},
  pages = {341},
  issn = {1044677X},
  doi = {10.6028/jres.114.025},
  urldate = {2025-07-28},
  langid = {english},
}

@incollection{hutchingsPointChargeCalculationsEnergy1964,
  title = {Point-{{Charge Calculations}} of {{Energy Levels}} of {{Magnetic Ions}} in {{Crystalline Electric Fields}}},
  booktitle = {Solid {{State Physics}}},
  author = {Hutchings, M.T.},
  year = {1964},
  volume = {16},
  pages = {227--273},
  publisher = {Elsevier},
  doi = {10.1016/S0081-1947(08)60517-2},
  urldate = {2025-07-28},
  isbn = {978-0-12-607716-2},
  langid = {english},
}

@article{kimSpinLiquidsAntiferromagnetic2013,
  title = {Spin {{Liquids}} and {{Antiferromagnetic Order}} in the {{Shastry-Sutherland-Lattice Compound}} \ce{Yb2Pt2Pb}},
  author = {Kim, M. S. and Aronson, M. C.},
  year = {2013},
  month = jan,
  journal = {Phys. Rev. Lett.},
  volume = {110},
  number = {1},
  pages = {017201},
  publisher = {American Physical Society},
  doi = {10.1103/PhysRevLett.110.017201},
  urldate = {2025-08-26},
}

@article{luoDesignHighEnergy2023a,
  title = {The {{Design}} of {{High Energy Direct Geometry Chopper Spectrometer}} at {{CSNS}}},
  author = {Luo, Wei and Feng, Yu and Liu, XinZhi and Wang, Meng and Zhu, DongHui and Gao, Wei and Geng, YanSheng and Ren, QingYong and Shen, JunYing and Sun, Yuan and Zhang, XiaoYue and Xia, YuanGuang and Zuo, TaiSen and Zheng, Yue and Tong, Xin},
  year = {2023},
  month = jan,
  journal = {Nucl. Instrum. Methods Phys. Res., Sect. A},
  volume = {1046},
  pages = {167676},
  issn = {0168-9002},
  doi = {10.1016/j.nima.2022.167676},
  urldate = {2025-09-09}
}

@article{gwph-w9zm,
  title = {Dipolar and quadrupolar correlations in the $5{d}^{2}$ {Re-based} double perovskites \ce{Ba2YReO6} and  \ce{Ba2ScReO6}},
  author = {Omar, Otkur and Zhang, Yang and Zhang, Qiang and Tian, Wei and Dagotto, Elbio and Chen, Gang and Arima, Taka-hisa and Stone, Matthew B. and Christianson, Andrew D. and Hirai, Daigorou and Gao, Shang},
  journal = {Phys. Rev. B},
  volume = {112},
  issue = {7},
  pages = {075103},
  numpages = {12},
  year = {2025},
  month = {Aug},
  publisher = {American Physical Society},
  doi = {10.1103/gwph-w9zm}
}

@book{Jensen:RareEarth,
    author = {Jensen, Jens and Mackintosh, Allan R},
    title = {Rare {{Earth Magnetism}}: {{Structures}} and {{Excitations}}},
    publisher = {Oxford University Press},
    year = {1991},
    month = {06},
    isbn = {9780198520276},
    doi = {10.1093/oso/9780198520276.001.0001}
}

@article{Rotter_2012,
doi = {10.1088/0953-8984/24/21/213201},
year = {2012},
month = {may},
publisher = {IOP Publishing},
volume = {24},
number = {21},
pages = {213201},
author = {Martin Rotter and Manh Duc Le and Andrew T Boothroyd and Jesus Angel Blanco},
title = {Dynamical matrix diagonalization for the calculation of dispersive excitations},
journal = {J. Phys.: Condens. Matter}
}

@article{PhysRevB.106.054431,
  title = {Excitonic transverse and amplitude fluctuations in noncollinear and charge-ordered \ce{RbFe^{2+}Fe^{3+}F6}},
  author = {Lane, H. and Songvilay, M. and Ewings, R. A. and Stock, C.},
  journal = {Phys. Rev. B},
  volume = {106},
  issue = {5},
  pages = {054431},
  numpages = {15},
  year = {2022},
  month = {Aug},
  publisher = {American Physical Society},
  doi = {10.1103/PhysRevB.106.054431}
}

@article{FOUCS1996,
  title = {Focus: {{Project}} of a {{Space}} and {{Time Focussing Time-of-Flight Spectrometer}} for {{Cold Neutrons}} at the {{Spallation Source SINQ}} of the {{Paul Scherrer Institute}}},
  author = {J. Mesot and S. Janssen and L. Holitzner and R. Hempelmann},
  journal = {J. Neutron Res.},
  volume = {3},
  issue = {4},
  pages = {293-310},
  year = {1996},
  date = {1996-09-01},
  doi = {10.1080/1023816960820020},
}

@article{Pula_2025,
doi = {10.1088/1361-648X/ae0f6c},
year = {2025},
month = {oct},
publisher = {IOP Publishing},
volume = {37},
number = {42},
pages = {425802},
author = {Pula, M and Sharma, S and Gautreau, J and K P, Sajilesh and Kanigel, A and dela Cruz, C R and Dolling, T N and Clark, L and Luke, G M},
title = {Fractional {{Magnetization Plateaus}} in the {{Shastry–Sutherland Lattice Material}} \ce{Er2Be2GeO7}},
journal = {J. Phys.: Condens. Matter},
}

@article{roscildeScalableSpinSqueezing2024,
  title = {Scalable {{Spin Squeezing}} from {{Critical Slowing Down}} in {{Short-Range Interacting Systems}}},
  author = {Roscilde, Tommaso and Caleca, Filippo and Angelone, Adriano and Mezzacapo, Fabio},
  year = {2024},
  date = {2024-11-21},
  journal = {Phys. Rev. Lett.},
  shortjournal = {Phys. Rev. Lett.},
  volume = {133},
  number = {21},
  pages = {210401},
  issn = {0031-9007, 1079-7114},
  doi = {10.1103/PhysRevLett.133.210401},
  urldate = {2025-12-02},
  langid = {english},
}

@book{schottky,
  title = {The specific heat of matter at low temperatures},
  author = {A. Tari},
  year = {2003},
  Address = {London},
  publisher = {Imperial College Press},
  isbn = {1860943144},
}

@article{yadavObservationUnprecedentedFractional2025a,
  title = {Observation of {{Unprecedented Fractional Magnetization Plateaus}} in a {{New Shastry-Sutherland Ising Compound}}},
  author = {Yadav, Lalit and Rufino, Afonso and Bag, Rabindranath and Ennis, Matthew and Koziol, Jan Alexander and Cruz, Clarina Dela and Kolesnikov, Alexander I. and Garlea, V. Ovidiu and Taddei, Keith M. and Graf, David and Schmidt, Kai Phillip and Mila, Frédéric and Haravifard, Sara},
  year = {2025},
  date = {2025-12-08},
  journal = {Phys. Rev. X},
  volume = {15},
  number = {4},
  pages = {041045},
  issn = {2160-3308},
  doi = {10.1103/9ynf-xx1t},
  urldate = {2025-12-24},
  langid = {english},
}

@article{gppd,
title = {The general purpose powder diffractometer at {{CSNS}}},
journal = {Physica B: Condensed Matter},
volume = {551},
pages = {370-372},
year = {2018},
issn = {0921-4526},
doi = {10.1016/j.physb.2017.11.005},
author = {Jie Chen and Le Kang and Huaile Lu and Ping Luo and Fangwei Wang and Lunhua He},
}

@article{ran_spin_2017,
  title = {Spin-Wave {{Excitations Evidencing}} the {{Kitaev Interaction}} in {{Single Crystalline}} $\alpha$-\ce{RuCl3}},
  author = {Ran, Kejing and Wang, Jinghui and Wang, Wei and Dong, Zhao-Yang and Ren, Xiao and Bao, Song and Li, Shichao and Ma, Zhen and Gan, Yuan and Zhang, Youtian and Park, J. T. and Deng, Guochu and Danilkin, S. and Yu, Shun-Li and Li, Jian-Xin and Wen, Jinsheng},
  journal = {Phys. Rev. Lett.},
  volume = {118},
  issue = {10},
  pages = {107203},
  numpages = {6},
  year = {2017},
  month = {Mar},
  publisher = {American Physical Society},
  doi = {10.1103/PhysRevLett.118.107203}
}

@article{yao_excitations_2022,
  title = {Excitations in the {{Ordered}} and {{Paramagnetic States}} of {{Honeycomb Magnet}} \ce{Na2Co2TeO6}},
  author = {Yao, Weiliang and Iida, Kazuki and Kamazawa, Kazuya and Li, Yuan},
  journal = {Phys. Rev. Lett.},
  volume = {129},
  issue = {14},
  pages = {147202},
  numpages = {7},
  year = {2022},
  month = {Sep},
  publisher = {American Physical Society},
  doi = {10.1103/PhysRevLett.129.147202}
}

@article{samarakoon_comparative_2017,
  title = {Comprehensive study of the dynamics of a classical {{Kitaev}} spin liquid},
  author = {Samarakoon, A. M. and Banerjee, A. and Zhang, S.-S. and Kamiya, Y. and Nagler, S. E. and Tennant, D. A. and Lee, S.-H. and Batista, C. D.},
  journal = {Phys. Rev. B},
  volume = {96},
  issue = {13},
  pages = {134408},
  numpages = {12},
  year = {2017},
  month = {Oct},
  publisher = {American Physical Society},
  doi = {10.1103/PhysRevB.96.134408}
}

@article{dorier_theory_2008,
  title = {Theory of Magnetization Plateaux in the {Shastry-Sutherland} Model},
  author = {Dorier, J. and Schmidt, K. P. and Mila, F.},
  journal = {Phys. Rev. Lett.},
  volume = {101},
  issue = {25},
  pages = {250402},
  numpages = {4},
  year = {2008},
  month = {Dec},
  publisher = {American Physical Society},
  doi = {10.1103/PhysRevLett.101.250402}
}

@article{corboz_crystals_2014,
  title = {Crystals of Bound States in the Magnetization Plateaus of the Shastry-Sutherland Model},
  author = {Corboz, Philippe and Mila, Fr\'ed\'eric},
  journal = {Phys. Rev. Lett.},
  volume = {112},
  issue = {14},
  pages = {147203},
  numpages = {5},
  year = {2014},
  month = {Apr},
  publisher = {American Physical Society},
  doi = {10.1103/PhysRevLett.112.147203}
}

@article{abendschein_effective_2008,
  title = {Effective Theory of Magnetization Plateaux in the Shastry-Sutherland Lattice},
  author = {Abendschein, A. and Capponi, S.},
  journal = {Phys. Rev. Lett.},
  volume = {101},
  issue = {22},
  pages = {227201},
  numpages = {4},
  year = {2008},
  month = {Nov},
  publisher = {American Physical Society},
  doi = {10.1103/PhysRevLett.101.227201}
}

@misc{buessen_spinmc_2020,
  title = {{SpinMC.jl}},
  year = {2020},
  howpublished = {\url{https://juliapackages.com/p/spinmc}},
}

@article{huang_multi_2012,
doi = {10.1088/0953-8984/24/38/386003},
year = {2012},
month = {aug},
publisher = {IOP Publishing},
volume = {24},
number = {38},
pages = {386003},
author = {Huang, W C and Huo, L and Tian, G and Qian, H R and Gao, X S and Qin, M H and Liu, J-M},
title = {Multi-step magnetization of the Ising model on a Shastry–Sutherland lattice: a Monte Carlo simulation},
journal = {J. Phys.: Condens. Matter}
}

@article{dublenych_ground_2012,
  title = {Ground States of the Ising Model on the Shastry-Sutherland Lattice and the Origin of the Fractional Magnetization Plateaus in Rare-Earth-Metal Tetraborides},
  author = {Dublenych, Yu. I.},
  journal = {Phys. Rev. Lett.},
  volume = {109},
  issue = {16},
  pages = {167202},
  numpages = {5},
  year = {2012},
  month = {Oct},
  publisher = {American Physical Society},
  doi = {10.1103/PhysRevLett.109.167202}
}

\end{document}